\documentclass[11pt,letterpaper]{book}

\usepackage{graphicx}
\usepackage{hyperref}
\usepackage{ifthen}

\usepackage{palatino}

\newcounter{refer}
\newcounter{shift}
\begin{document}

\setcounter{page}{-1}


\setlength{\unitlength}{1mm}
\begin{picture}(0,0)
\put(-5,0){\includegraphics[width=230mm,angle=-90]{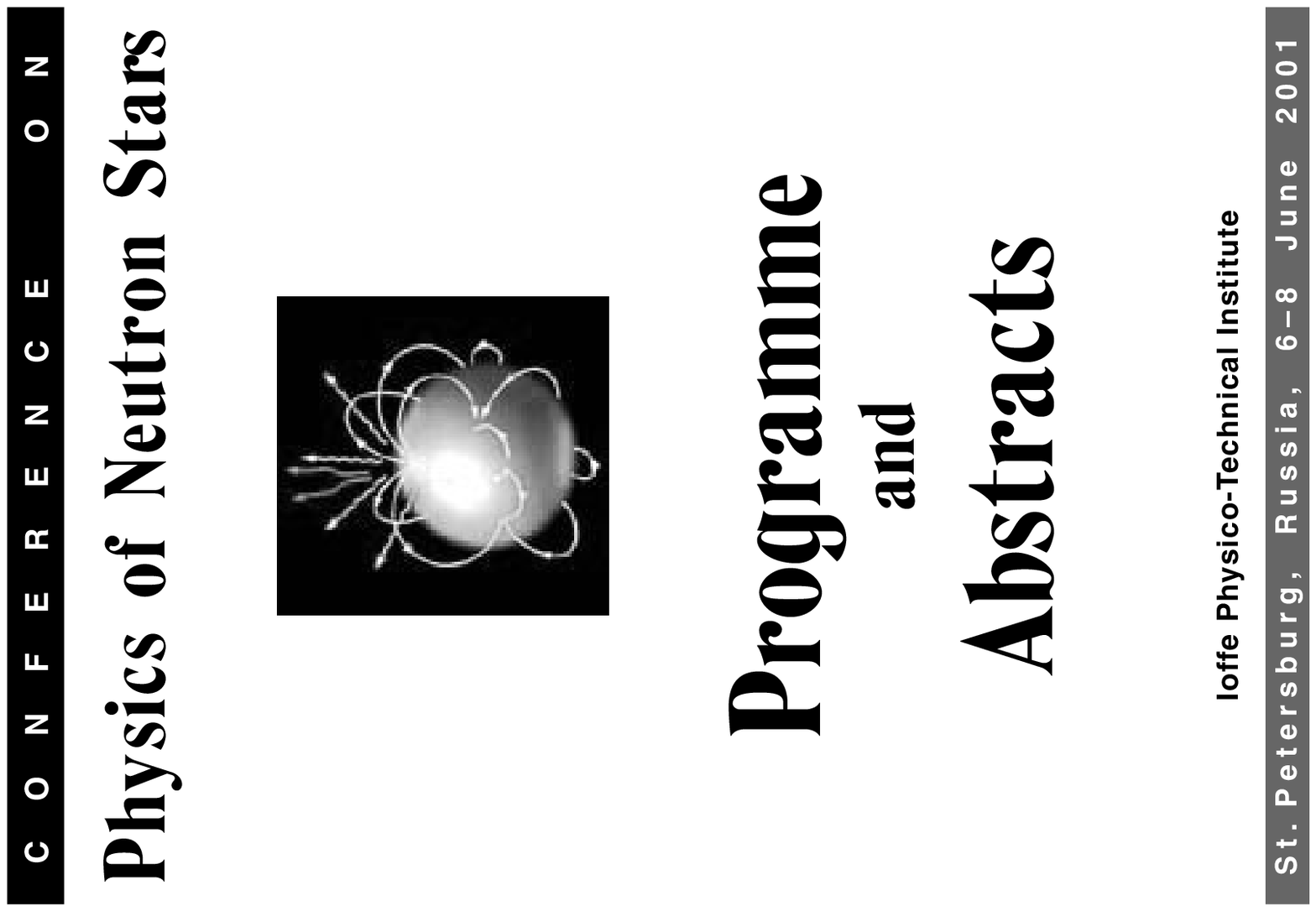}}
\end{picture}

\thispagestyle{empty}

\newpage
~
\thispagestyle{empty}
\newpage
~
\vspace{1cm}
\begin{center}
\Huge\bf
Physics of Neutron Stars \\
2001
\end{center}
\thispagestyle{empty}

\vspace{1cm}

The Conference was held at the A.F.~Ioffe
Physical-Technical Institute (Ioffe Inst.) 
since June 6 to June 8, 2001.
This was the sixth gathering on neutron star physics
in St.-Petersburg (after those in 1988, 1992,
1995, 1997, and 1999). Its aim was to bring together
scientists interested in physics and astrophysics of neutron stars.
All subjects relevant to theory and observations of neutron stars were welcome.

The event gathered about 70 participants from 
15 Russian institutions of St.-Petersburg
(Ioffe Institute, 
Pulkovo Observatory,
St.-Petersburg State University),
Moscow
(Sternberg Astronomical Institute, 
Moscow State University,
Institute for Experimental and Theoretical Physics, 
Lebedev Physical Institute, 
Astro Space Center of the Russian Academy of Sciences, 
Astrophysics Institute at the Moscow State Engineering Physics Institute,
Institute for Nuclear Research,
Space Research Institute,
Pushchino Radio Astronomical Observatory), 
Nizhny Novgorod (Institute of Applied Physics), 
Yaroslavl (Yaroslavl State University), and
Nizhny Arkhyz (Special Astrophysical Observatory).
There were also three foreign participants from
Copernicus Astronomical Center (Poland) and
George Mason University / Naval Research Laboratory (USA).

The three day meeting consisted of 12 sessions. 
Their main subjects were:
radio pulsars;
neutron stars as X-ray sources;
internal structure and evolution of neutron stars;
observations of isolated neutron stars;
disks, QPOs, magnitospheres;
neutron stars in the Galaxy and beyond;
neutrino processes;
gamma-ray bursts.
The sessions included 9 review talks of 30--40 min.\
duration and 40 contributed talks (20 min.).
In addition, there were 8 poster contributions.

The abstracts and programme were published in
the form of a booklet "Programme and Abstracts"
(Ioffe Physical Technical Institute, 49 pages, 2001)
before the Conference. 
Two changes appeared in the programme during the Conference:
the talk of V.M.~Malofeev and O.I.~Malov (see p.~\pageref{malov2}) was not presented
and the contribution of A.G.~Kuranov, K.A.~Postnov and M.E.~Prokhorov
(see p.~\pageref{kuranov})
was presented as a poster.

The booklet is presented here; it can 
also be accessed at the Ioffe Institute web pages:
\href{http://www.ioffe.rssi.ru/astro/NS2001/index.html}
{\bf http:/$\!$/www.ioffe.rssi.ru/astro/NS2001/index.html}.

A part of abstracts contain references to the web pages
where the reader may find additional information on presentations.
A part of abstracts contain references to published articles.

Conference was supported by the Russian
Foundation for Basic Research,
grant 01-02-26057, plus additional support for young scientists.
A help of administration
of the Ioffe Institute is also acknowledged. 
It was generally agreed by participants that 
Neutron Stars Conference was a useful and productive 
scientific event; it would be desirable to organize such
meetings regularly, presumably once in two years.

\newpage


\thispagestyle{empty}







\begin{center}
\Large
The Conference is organized by \\
{\bf the Ioffe Physical Technical Institute }
\end{center}

\vspace*{1cm}\hspace*{2cm}

{\sl The Organizing Committee}:

\begin{list}{$\diamond$}%
            {\usecounter{refer}
             \leftmargin 15mm
             \itemsep 0mm}

  \item
  {\bf D.A.~Varshalovich} ({\it Chair,\/} Ioffe Physical Technical Institute, St.~Petersburg)

  \item
  R.L.~Aptekar (Ioffe Physical Technical Institute)

  \item
  G.S.~Bisnovatyi-Kogan (Space Research Institute, Moscow)

  \item
  Yu.N.~Gnedin (Pulkovo Astronomical Observatory, St.~Petersburg)

  \item
  A.V.~Ivanchik (Ioffe Physical Technical Institute)

  \item
  A.D.~Ka\-min\-ker (Ioffe Physical Technical Institute)

  \item
  A.B.~Koptsevich (Ioffe Physical Technical Institute)

  \item
  K.A.~Postnov (Sternberg Astronomical Institute)

  \item
  A.Y.~Potekhin (Ioffe Physical Technical Institute)

  \item
  M.E.~Prokhorov (Sternberg Astronomical Institute)

  \item
  Yu.A.~Shibanov (Ioffe Physical Technical Institute)

  \item
  D.G.~Yakovlev (Ioffe Physical Technical Institute)

\end{list}

\vspace{1cm}

The Conference is held at the Big Hall of the Main Building of the A.F.~Ioffe 

Physical-Technical Institute for three days, from June 6 to June 8, 2001.\\

{\sl Address}:

Politekhnicheskaya 26, St.-Petersburg 194021 (metro station Politekhnicheskaya)

ph. (812) 2479180, 2479326; local ph. 180, 326

\vspace{1cm}

\href{http://www.ioffe.rssi.ru/astro}{{\bf http://www.ioffe.rssi.ru/astro}}

\vspace{1cm}



\vspace{1cm}



\thispagestyle{empty}

\vfill


\copyright\ Ioffe Physical Technical Institute, St. Petersburg, 2001


\newcommand{\speaker}[1]{\underline{#1}}

\newcommand{\nl}{}

\newcommand{\nli}{}

\newcommand{\inst}[1]{}

\newcommand{\instaddr}[1]{}

\newcommand{\contribution}[9]{#2 {\it (#3)}. #1}

\newcommand{\abstext}[1]{}

\newcommand{\arxiv}[1]{#1}

\newcommand{\references}[1]{}


\newpage\normalsize
~

\markboth{\sl Programme}
         {\sl Physics of Neutron Stars -- 2001}


\section[Programme of the Conference]{}

\vspace*{8cm}
\noindent
\centerline{\Huge\bf PROGRAMME}
\thispagestyle{empty}
\newpage
~
\thispagestyle{empty}
\newpage


\begin{center}
\large\bf June 6, Wednesday
\end{center}

\begin{list}{10.00--10.15~}
            {
             \itemsep -0.1mm}

\item[{\it Session 1.}] 
{\sl WELCOME and GENERAL}\\
Chairman: D.A.~Varshalovich

\item[\bf 10.00--10.10] 
D.A.~Varshalovich {\it(Ioffe Institute)}. Welcome

\item[\bf 10.10--10.50]
\contribution
%
%
{Phase transitions in stars: stability, pulsations, \nl
convective Urca-shells and pre-supernovae}
%
%
%
{\speaker{G.S.~Bisnovatyi-Kogan}\inst{1}}
{%
\inst{1}%
Space Research Institute%
\instaddr{, 84/32, Profsoyuznaja st., 117810, Moscow, Russia \\}}
{gkogan@mx.iki.rssi.ru}
%
%
%
{}
{}
{astro-ph/0004281}{}{}


\abstext
{
The problem of damping of stellar oscillations in the presence of Urca
shell is solved analytically in a plane symmetrical approximation.
Low amplitude oscillations are considered. Oscillatory pressure
perturbations induce beta reactions of the electron capture and
decay in a thin layer around the Urca shell, leading to damping
of oscillations. Due to nonlinear dependence of beta reaction
rates on the pulsation amplitude in a degenerate matter, even a
low amplitude oscillation damping follows a power-low. It is shown
that in the presence of the Urca shell the energy losses due to
neutrino emission, and the entropy increase due to non-equilibrium
beta reactions are much lower than the rate of decrease of the
energy of pulsations by excitation of short-wavelength
acoustic waves. Dissipation of the vibrational energy by the
latter process is the main source of heating of matter. Convective
motion in the presence of the Urca shell is considered, and
equations generalizing the mean free path model of convection
are derived. Convective motion is a source of both energy losses 
due to Urca reactions
in the shell, and nonequilibrium beta heating of degenerate matter.
This problem is closely related to thermal stability 
and boundary of SN type I explosions.
Only self-consistent evolutionary calculations may 
clarify the effect
of convective Urca-shell on the thermal stability 
of the pre-SN model.
}

%
%

\item[\bf 10.50--11.10]
\contribution
%
%
{From white dwarfs to black holes \nl
(70th anniversary of the theory of compact objects)}
%
%
%
{\speaker{A.F.~Zakharov}\inst{1}}
{%
\inst{1}%
Institute for Theoretical and Experimental Physics%
\instaddr{, 117259, Moscow, Russia \\}}
{zakharov@vitep5.itep.ru}
%
%
%
{}
{}
{}{}{}


\abstext
{
We discuss basic ideas which laid foundation of the black hole
concept. The major goal of the historical part is to
explain the very long way of the birth of the black hole concept.
The black hole solution was derived by K.~Schwarzschild
in 1916, but black hole concept was introduced by J.A.~Wheeler
only in 1967. We emphasize the great contribution of S.~Chandrasekhar
into development of this concept.
We discuss the basic notations of the black hole theory
and observational (astronomical) manifestations of black holes,
for example, we analyse a possibility to interpret 
the very peculiar distortion of the iron K$_{\alpha}$-line 
in such a way.
}

%
%

\item[\bf 11.10--11.30]
\contribution
%
%
{r-process in neutron star mergers and beta-delayed fission }
%
%
%
{\speaker{I.V.~Panov}\inst{1}, F.-K.~Thielemann\inst{2}}
{%
\inst{1}%
Institute for Theoretical and Experimental Physics,%
\instaddr{ Moscow, 117259, Russia \\}
\inst{2}%
University of Basel%
\instaddr{, Klingelbergstr. 82, CH-4056 Basel, Switzerland \\}}
{panov@mpa-garching.mpg.de, panov@vitep5.itep.ru}
%
%
{http://www.itep.ru/lab230/ns2001.html}
{http:/$\!$/www.itep.ru/lab230/ns2001.html}
{}{}{}


\abstext
{The studies of beta-delayed fission in r-process have a long
history, but they have mainly been focused on cosmochronometers. 
It has been thought that its effect has small influence on the
majority of r-process products, taken into account realistic fission
barriers.
For that reason the majority of investigations of the r-process
employed
beta-delayed fission of transuranium nuclei in a very simplified manner:
$P_{\beta df} \equiv P_{sf}=1$ for all A $>$ $A_{fiss}$ 
(see [1] and references therein).

That is why reliability 
of both, beta-delayed fission rates 
(first of all, for cosmochronology) and 
mass distribution of fission products
(important mainly for formation of nuclei with mass numbers A $<$ 130
[2]), has not been studied.
In this work we make an attempt to solve these problems numerically.     
The kinetic network [3]  calculations
of r-process for conditions in neutron star mergers  
[4] have been performed. Different theoretical beta-delayed 
fission probabilities have been used, and different 
mass distributions of fission products have been considered. 

Calculations with different fission rates  show strong 
dependence on theoretical physics input. 
Our calculations  give better agreement with
observations, first of all in relative yields for peaks $A\approx
$130, 196 and for nuclei with $ Z \approx$ 44--48. 
The problem of realistic mass distribution of nuclear fission
products is still open because of poor knowledge of fission  
of very neutron rich transuranium nuclei. 
However there are some indications
from both, nuclear physics and astrophysical observations [5], 
that asymmetric fission should be the main fission mechanism. 
The masses
of fission products have to be 
determined taking into account  shell effects. 
Preliminary results show significant dependence of the yields 
of nuclei-cosmochronometers from different fission data involved in
calculations. 
Comparison of the results with observations 
of r-elements in very metal poor stars 
may help to find limits on  
probable mass distribution of fission products.
}

\references{

\item
 J.J.~Cowan, B.~Pfeiffer, Kratz K.-L. et al., ApJ 521, 194 (1999) 

\item  I.V.~Panov, C.~Freiburghaus, F.-K.~Thielemann.
In: Proc. of 10 Workshop on Nucl. Astroph., Ringberg, 2000,
eds. W.~Hillebrand, E.~M\"oller. MPA/P12, P.73

\item I.V.~Panov, S.I.~Blinnikov, F.-K.~Thielemann,
 Astronomy Letters.  27, 279 (2001) 

\item C.~Freiburghaus, S.~Rosswog, F.-K.~Thielemann.
  ApJ  525, L121 (1999) 

\item C.~Sneden, J.J.~Cowan, I.I.~Ivans et al.
 ApJ  533, L139 (2000)

}

\medskip
\item[\bf 11.30--12.00]
{\it Coffee break}
\bigskip

\item[\bf {\it Session 2.}]
{\it RADIOPULSARS}\\
Chairman: M.E.~Prokhorov

\item[\bf 12.00--12.20]
\contribution
%
%
{Comparative analysis of radio luminosity \nl
of millisecond and normal pulsars }
%
%
{\speaker{A.D.~Kuzmin}\inst{1}}
{%
\inst{1}%
Pushchino Radio Astronomy Observatory%
\instaddr{, Pushchino, 142292, Moscow Region, Russia \\}}
{akuzmin@prao.psn.ru}%
%
%
{}
{}
{}{}{}


\abstext {
We present the results of 
comparative analysis of
the integral radio luminosity
of the millisecond and normal pulsars.

Analysis is based on our measurements of the flux densities,
spectra and integral luminosities of 30 millisecond pulsars and
on the data, borrowed from the literature, which allows us to construct
the integral radio luminosities of 485 ``normal'' pulsars.

We find that contrary to the great difference of millisecond and
``normal'' pulsars in spin periods $P$, period derivatives $\dot P$,
magnetic field strengths $B$, and characteristic ages $\tau$, the
integral radio luminosities of these two pulsar populations are
nearly equal. The same is true for their dependences
on $P, \dot P, B, \tau$ and on losses of the kinetic energy $\dot E$.

We find that the integral luminosities of both millisecond and
``normal'' populations of pulsars are nearly proportional to the
parameter $B/P^2$, which characterizes potential difference
between the base and the top of the gap of a polar cap.

We suggest that millisecond and ``normal'' pulsars have similar
mechanism of radio emission, in which energetic properties are
controlled by the potential difference between the base and the
top of the gap of the polar cap.}

%
\references{

\item
A.D. Kuzmin, B.Ya. Losovsky. A\&A 368, 230 (2001)

\item
J.H. Taylor, R.N. Manchester, A.G. Lyne, F. Camilo, unpublished
 (1995)
}

\item[\bf 12.20--12.40]
\contribution
%
%
{On the pulsed optical emission from radio pulsars}
%
%
%
{\speaker{I.F.~Malov}\inst{1}}
{%
\inst{1}%
Pushchino Radio Astronomy Observatory%
\instaddr{, Pushchino, 142292, Moscow Region, Russia \\}}
{malov@prao.psn.ru}
%
%
%
{}
{}
{}{}{}


\abstext
{
The formula for a radio pulsar luminosity associated 
with synchrotron emission of the primary beam is obtained. 
The formula is based on the model of an emitting torus at 
the light cylinder and on the solution of the kinetic equation for 
pitch-angle distribution of relativistic particles. The 
high correlation between the observed optical luminosity of radio 
pulsars and the parameter $\dot P / P^4$ is found: 
$\log \left( \frac{L_{opt}}{L_{Crab}} \right)$ = (1.30 $\pm$ 0.19) 
$\log{\dot P_{-14} / P^4} - 4.21 \pm 1.02$ (the correlation coefficient 
$\rho$ = 0.97 $\pm$ 0.14). Here $P$ is the pulsar period, $\dot P$ 
is its derivative. This correlation 
allows one to predict  
possible optical emission from several dozens of pulsars 
(in particularly, from all pulsars with $P < 0.1$ sec). 
Comparison of this prediction with multiwavelength
observations of radio pulsars shows that the predicted 
list contains all 27 known emitters of hard radiation. The shift 
of maximum frequency in the synchrotron spectrum to higher 
frequencies with decreasing period $P$ is predicted. This 
prediction is in agreement with data for the same 27 pulsars. 
The obtained results show that the synchrotron model describes 
the main properties of non-thermal optical and harder emission 
of radio pulsars.
}

%
%

\item[\bf 12.40--13.00]
\contribution
%
%
{First detection of pulsed radio emission from an AXP}
%
%
%
{V.M.~Malofeev\inst{1}, \speaker{O.I.~Malov}\inst{1}}
{%
\inst{1}%
Pushchino Radio Astronomy Observatory%
\instaddr{, Pushchino, 142292, Moscow Region, Russia \\}}
{malofeev@prao.psn.ru}
%
%
%
{}
{}
{}{}{}


\abstext
{
We report on the discovery and some investigations 
of pulsed radio emission with period 6.98 s from the anomalous 
X-ray pulsar (AXP) 1E 2259+586. The observations were made from 
March 1999 to April 2001 at 111.5 MHz with the Large Phased Array 
in Pushchino. 
The mean flux density is about 70 mJy, the integrated profile is narrow 
and its width at 50 percent of maximum intensity is approximately 
120 ms. The dispersion measure is 80 $\pm$ 5 pc cm$^{-3}$ that gives 
the distance to the pulsar of about 3.6 kpc. This value is confirmed with 
the estimation of the distance to SNR G109.1$-$1.0 (3.6 -- 4.7 kpc) 
with the X-ray - bright central pulsar 1E 2259+586. 
}

%
%

\item[\bf 13.00--13.20]
\contribution
%
%
{Observed parameters of microstructure \nl in pulsar radio emission}
%
%
%
{\speaker{M.V.~Popov}\inst{1}, V.I.~Kondratev\inst{1}}
{%
\inst{1}%
Astro Space Centre FIAN%
\instaddr{, 117810, Moscow, Russia \\}}
{mpopov@asc.rssi.ru}
%
%
%
{}
{}
{}{}{}


\abstext
{
The microstructure of several bright pulsars was 
investigated with a time resolution of 62.5 ns. The pulsars 
were observed with the 70-m NASA/DSN radio telescope at Tidbinbilla, 
Australia, at a frequency of 1650 MHz. Histograms of microstructure 
time scales show steep increase toward shorter time scales followed 
by a sharp cutoff at about $5-10$ $\mu$s. The shortest micropulse 
detected has a width of 2 $\mu$s.
No unresolved nanopulses or pulse structure with submicrosecond 
time scale were found. The statistics of the micropulses and 
their quasi-periodicities differ significantly for two 
components of PSR B1133+16. Microstructure quasi-periodicities 
are most likely unrelated to any modes of vibrations of neutron stars.
}

%
%

\item[\bf 13.20--13.40]
\contribution
%
%
{Diffractive scintillations of PSR 0809+74 and \nl
PSR 0950+08 at low frequencies}
%
%
%
{\speaker{T.V.~Smirnova}\inst{1}}
{%
\inst{1}%
Pushchino Radio Astronomy Observatory%
\instaddr{, Pushchino, 142292, Moscow Region, Russia \\}}
{tania@prao.psn.ru}
%
%
%
{}
{}
{}{}{}


\abstext
{
Low frequency individual pulse observations were 
carried out for the pulsars PSR 0809+74 and PSR 0950+08 in the range 
from 64 to 111 MHz. Frequency - time structure of  emission was 
studied to separate internal (due to pulsar) and external 
intensity variations due to electron density irregularities 
in the interstellar plasma. We report characteristic time 
and frequency scales of these variations at low frequencies.
}

%
%

\item[\bf 13.40--14.00]  
\contribution
%
%
{Giant pulses from radiopulsars}
%
%
%
{\speaker{V.A.~Soglasnov}\inst{1}, M.V.~Popov\inst{1}, V.I.~Kondratev\inst{1},
S.V.~Kostyuk\inst{1}}
{%
\inst{1}%
Astro Space Centre FIAN%
\instaddr{, 117810, Moscow, Russia \\}}
{vsoglasn@asc.rssi.ru}
%
%
%
{}
{}
{}{}{}


\abstext
{
There are two pulsars, where remarkable phenomenon of 
"giant pulses" is observed: the Crab pulsar and the 
millisecond pulsar B1937+21. We present new results of our high 
time resolution observations of giant pulses from these pulsars. 
The events with extremely high flux, 300 000 Jy (Crab) and 65 000 Jy 
(1937+21), were detected. Many properties of giant pulses become 
rather unexpected (for instance, a very short - nanosecond - 
duration of GP from 1937+21), they are important for understanding 
physics of neutron star magnetosphere.
}

%
%

\medskip
\item[\bf 14.00--15.00] 
{\it Lunch}
\bigskip

\item[{\it Session 3.}]
{\it NEUTRON STARS AS X-RAY SOURCES. Part 1} \\
Chairman: A.D.~Kuzmin

\item[\bf 15.00--15.30]
\contribution
%
%
{Short-term X-ray variability of neutron stars: \nl current status
and perspectives}
%
%
%
{\speaker{M.G.~Revnivtsev}\inst{1}, M.R.~Gilfanov\inst{1},
E.M.~Churazov\inst{1}, R.A.~Sunyaev\inst{1}}
{%
\inst{1}%
Space Research Institute%
\instaddr{, 84/32, Profsoyuznaja st., 117810, Moscow, Russia \\}}
{revnivtsev@hea.iki.rssi.ru}
%
%
%
{}
{}
{}{}{}


\abstext
{
The black holes and neutron stars are the fastest objects in the
Universe. From the very beginning of X-ray astronomy there was a
search for the shortest timescales in the X-ray variability of 
these compact objects. The great recent improvement in this topic was made
with the help of Rossi X-ray Timing Explorer observatory. In this talk 
I would like to describe our latest results on the high frequency continuum
variability of black holes and neutron stars. A special attention
will be paid to the different patterns of X-ray variability of black holes
and neutron stars at hundreds of Hz and to the continuum noise of
compact objects at the highest available frequencies -- up to $\sim$30 kHz. 
}

%
%

\item[\bf 15.30--15.50]
\contribution
%
%
{X-ray bursts with signs of strong photospheric radius \nl
expansion of a neutron star}
%
%
%
{\speaker{S.A.~Grebenev}\inst{1}, S.V.~Molkov\inst{1}, 
A.A. Lutovinov\inst{1}, R.A. Sunyaev\inst{1}}
{%
\inst{1}%
Space Research Institute%
\instaddr{, 84/32, Profsoyuznaja st., 117810, Moscow, Russia \\}}
{sergei@hea.iki.rssi.ru}
%
%
%
{}
{}
{astro-ph/0005082}{}{}


\abstext
{
We present results of the GRANAT/ART-P and RXTE/PCA observations of
several X-ray bursts with obvious signs of strong photospheric radius
expansion (precursors and dips in profiles, delayed rise phase, etc).
In particular, the photospheric radius exceeded 70 km during an
intense X-ray burst detected from the source 1E1724-307 in the
globular cluster Terzan 2. After the precursor which was comparable
in strength with the main event the flux from the source fell below
the persistent level. The very narrow precursor was detected in the
profile of a long ($\sim$ 10 min) X-ray burst observed from GX17+2. We
analyze correlations in the effective temperature and radius of the
photosphere in these and other X-ray bursters and note that
Comptonization may be responsible for the detected features. To
describe in detail spectral evolution of the source during the burst
a number of model spectra were computed taking into account both
Comptonization and free-free absorption in the outer layers of the
photosphere.
}

%
%

\item[\bf 15.50--16.10]
\contribution
%
%
{Review of X-ray bursters in the Galactic Center Region}
%
%
%
{\speaker{A.A.~Lutovinov}\inst{1}, S.A.~Grebenev\inst{1}, S.V.~Molkov\inst{1}}
{%
\inst{1}%
Space Research Institute%
\instaddr{, 84/32, Profsoyuznaja st., 117810, Moscow, Russia \\}}
{aal@hea.iki.rssi.ru}
%
%
%
{}
{}
{astro-ph/0009349}{astro-ph/0105001}{astro-ph/0105002}


\abstext
{
Results of observations of X-ray bursters in the Galactic Center
region carried out with the RXTE observatory and the ART-P telescope
on board GRANAT are presented. Eight X-ray bursters (A1742-294,
SLX1744-299/300, GX3+1, GX354-0, SLX1732-304, 4U1724-307, KS1731-260)
were studied in this region during five series of observations which
were performed with the ART-P telescope in 1990-1992 and more than 100 
type I X-ray bursts from these sources were observed. For each of the
sources we investigated in detail the recurrence times between bursts,
the bursts time profiles and their dependence on the bursts flux, the
spectral evolution of source emission in the persistent state and during
bursts. Two bursters (SLX1732-304, 4U1724-307) located in the
globular clusters Terzan 1 and 2 were investigated also using the RXTE 
data.
}

%
%

\item[\bf 16.10--16.30]
\contribution
%
%
{X-ray spectral variability of the LMXB and \nl Z-source GX340+0}
%
%
%
{\speaker{S.V.~Molkov}\inst{1}, S.A.~Grebenev\inst{1}, A.A.~Lutovinov\inst{1}}
{%
\inst{1}%
Space Research Institute%
\instaddr{, 84/32, Profsoyuznaja st., 117810, Moscow, Russia \\}}
{molkov@hea.iki.rssi.ru}
%
%
%
{}
{}
{}{}{}


\abstext
{
We present the results of analysis of the PCA/RXTE data obtained 
during long ($\sim$ 390 ks) pointing towards the Z-source GX~340+0. 
The complete Z track has been traced out for this source in 
the color-color diagram. For the analysis we separated the whole 
set of data (24 separate observations) in 16-s time segments and 
accumulated X-ray spectra for each of the segments. We studied 
spectral behaviour of the source as a function of its position 
in the color-color diagram. In general the spectra could not be 
fitted with any simple model. We used two-component model: 
bremsstrahlung plus black-body with fixed photoelectric absorption. 
Our analysis reveals that the role of black-body emission increases 
during transition from the horizontal branch of Z track to the 
flared branch.
}

%
%

\medskip
\item[\bf 16.30--17.00]
{\it Coffee break}
\bigskip

\item[{\it Session 4.}]
{\it NEUTRON STARS AS X-RAY SOURCES. Part 2}\\
Chairman: S.A.~Grebenev

\item[\bf 17.00--17.20]
\contribution
%
%
{ Hard-energy emission from supernova remnants}
%
%
%
{\speaker{A.M.~Bykov}\inst{1}}
{%
\inst{1}%
Ioffe Physical Technical Institute%
\instaddr{, Politekhnicheskaya 26, St. Petersburg 194021 \\}}
{byk@astro.ioffe.rssi.ru }
%
%
%
{http://www.ioffe.rssi.ru/astro/DTA/DTA-Pub1.html}
{http:/$\!$/www.ioffe.rssi.ru/astro/DTA/DTA-Pub1.html}
{astro-ph/0010157}{}{}


\abstext
{Hard X-ray and gamma-ray emission has been detected from a number of 
galactic supernova remnants. We discuss relevant processes and models of 
hard emission production in supernova remnants simultaneously with the results 
of recent observations of SNRs obtained with CGRO, ASCA, BeppoSAX and 
XMM-Newton. Implications of the observed spectra of SNRs to test models of 
pulsar wind nebulae emission as well as the cosmic ray origin problem will 
be discussed.       
}

%
%
%
%

\item[\bf 17.20--17.40]
\contribution
%
%
{A model of sub-Eddington accretion onto a magnetized \nl neutron star}
%
%
%
{\speaker{A.M.~Krassilchtchikov}\inst{1}, A.M.~Bykov\inst{1}}
{%
\inst{1}%
Ioffe Physical Technical Institute%
\instaddr{, Politekhnicheskaya 26, St. Petersburg 194021 \\}}
{kra@astro.ioffe.rssi.ru}
%
%
%
{}
{}
{}{}{}


\abstext
{
A non-stationary one-dimentional collisionless
two-fluid model of sub-Eddington accretion onto a
magnetized neutron star is developed to be used
for modeling of hard emission from X-ray binaries.
        Temporal evolution of accreting flows is
studied for a range of accretion rates and magnetic
fields within a first-order Godunov scheme with source
terms splitting.
        Strong shocks accompanied by hot plasma
regions are found to develop on timescales of about
10$^{-5}$ s; they are stable up to 10$^{-2}$ s or even longer.
        Hard emission from the hot regions may be
detected by modern X-ray and gamma-ray missions.
In this way parametres of the model may be constrained
and typical physical conditions in the flow revealed.
}

%
%

\item[\bf 17.40--18.00]
\contribution
%
%
{Theoretical interpretation of the X-ray pulsar 
spectra \nl consisting of several cyclotron harmonics}
%
%
%
{\speaker{A.N.~Baushev}\inst{1}, G.S.~Bisnovatyi-Kogan\inst{1}}
{%
\inst{1}%
Space Research Institute%
\instaddr{, 84/32, Profsoyuznaja st., 117810, Moscow, Russia \\}}
{abaushev@mx.iki.rssi.ru, gkogan@mx.iki.rssi.ru}
%
%
%
{}
{}
{}{}{}


\abstext
{
The spectrum of cyclotron radiation is calculated 
for anisotropically distributed relativistic electrons 
moving with  
nonrelativistic velocities across the magnetic field. 
It is shown that if such electrons
are responsible for the formation of the ``cyclotron'' line in 
the spectrum of Her X-1 then the value of its magnetic field, 
(3--6)$\cdot$ 10$^{10}$ G, which results from
this interpretation, is 
in a good agreement with some other observations and theoretical 
estimates. The case of multi-harmonical cyclotron line 
is also considered. Observations of temporal variations 
of the energy of this ``cyclotron'' line in the spectra of several 
X-ray pulsars is explained by variability of the average 
longitudinal energy of electrons, which decreases with increasing 
luminosity due to radiative deceleration of the accretion flow.
}

%
%

\item[\bf 18.00--18.20]
\contribution
%
%
{Cyclotron scattering with mode switching 
in accretion \nl column of a magnetized neutron star}
%
%
%
{\speaker{A.V.~Serber}\inst{1}}
{%
\inst{1}%
Institute of Applied Physics%
\instaddr{, 46 Ulyanov st., 603600, Nizhny Novgorod, Russia \\}}
{serber@appl.sci-nnov.ru}
%
%
%
{}
{}
{}{}{}


\abstext
{
We consider scattering of cyclotron radiation 
at the first harmonic in a rarefied plasma near a neutron 
star with a dipole magnetic field. It is assumed that the
case of strongly inhomogeneous magnetic field is realized,
i.e., the size of the plasma region is large compared 
to the size of gyroresonance layer in the neutron-star 
magnetic field. This case is analyzed under the 
conditions that normal-wave polarization at the first 
harmonic is determined by vacuum birefringence and the 
efficient switching of the linearly-polarized ordinary 
and extraordinary modes takes place due to the cyclotron 
scattering at the first harmonic. The obtained solution
 of the corresponding equation of cyclotron-radiation 
transfer reveals that the cyclotron scattering with mode 
switching leads to an efficient depolarization of
radiation outgoing from an optically thick 
gyroresonance layer. 
}

%
%

\item[\bf 18.20--18.40]
\contribution
%
%
{X-ray emission from young supernova remnants}
%
%
%
{\speaker{D.I.~Kosenko}\inst{1}, S.I.~Blinnikov\inst{1,2},
E.I.~Sorokina\inst{1}, K.A.~Postnov\inst{1}}
{%
\inst{1}%
Sternberg Astronomical Institute,%
\instaddr{ Universitetskij pr. 13, 119899, Moscow, Russia \\} 
\inst{2}%
Institute for Theoretical and Experimental Physics%
\instaddr{, 117259, Moscow, Russia \\}}
{lisett@xray.sai.msu.ru}
%
%
%
{}
{}
{}{}{}


\abstext
{
We present the results of 
hydrodynamical simulations of the Tycho SN remnant. Our
model is one-dimensional and spherically symmetrical but
it takes into account
kinetics and various types of radiative transfer. 
We can obtain detailed theoretical X-ray luminosity
profiles of the remnant in different lines (silicon and iron) and
compare it with the observational results. There are several models of
type Ia SNe with different abundances and density profiles, so we could make an
assumption about progenitor of the Tiho SN and choose the most viable
model.
}

%
%


\bigskip
\begin{center}
\large\bf June 7, Thursday
\end{center}
\medskip

\item[{\it Session 5.}]
{\it INTERNAL STRUCTURE AND EVOLUTION}\\
Chairman: I.F.~Malov

\item[\bf 10.00--10.40]
\contribution
%
%
{Unified equations of state of dense matter and \nl neutron star structure }
%
%
%
{\speaker{P. Haensel}\inst{1}}
{%
\inst{1}%
N. Copernicus Astronomical Center%
\instaddr{, Bartycka 18, 00-716 Warszawa, Poland\\}}
{haensel@camk.edu.pl}
%
%
%
{}
{}
{astro-ph/0006135}{astro-ph/0105485}{}


\abstext
{
Unified equation of state (EOS)  describes in a physically 
consistent way neutron star matter within the crust and the liquid 
core, and is based on a single nuclear hamiltonian, ${\hat H}_{\rm N}$. 
To make a solution of the nuclear many-body problem feasible, one uses 
an effective nuclear hamiltonian, ${\hat H}^{\rm eff}_{\rm N}$. It has 
to reproduce the ground state properties of laboratory nuclei, in
particular of those with a large neutron excess, and to 
give correct saturation 
parameters of nuclear matter. Additional conditions, relevant for correct
description of very neutron-rich matter in neutron stars have to be also 
imposed on ${\hat H}_{\rm N}^{\rm eff}$. 
Two examples of  ${\hat H}_{\rm N}^{\rm eff}$ are: an older 
FPS ({\bf F}riedman-{\bf P}andharipande-{\bf S}kyrme) 
 model (Pandharipande \& Ravenhall 1989), and a recent SLy 
({\bf S}kyrme-{\bf L}yon) one 
(Chabanat et al. 1997, 1998). Corresponding unified EOSs were  
calculated, assuming ground state, $T$=0, for  the crust matter, 
and  simplest (minimal) $npe\mu$ composition of the 
liquid core: FPS EOS (Lorenz et al. 1993), SLy EOS (Douchin \& Haensel 2000, 
2001). 
 Properties of the bottom layers of the neutron star 
crust will be  discussed, with particular emphasis on the crust core 
interface (Douchin et al., Douchin \& Haensel 2000). 
 Neutron star models, calculated using  these EOSs (Douchin \& Haensel 2001), 
 will be  reviewed. In particular, properties of 
equilibrium configurations  near minimum mass  (Haensel, Zdunik, 
Douchin 2001), will be  studied, and related 
to the specific features of the EOS near crust-core interface.  
 Effects 
of rotation on both $M_{\rm max}$ and $M_{\rm min}$, and on the 
mass-radius curves  
(Haensel, Zdunik, Douchin 2001) 
 will be 
briefly  reviewed.  
Finally, parameters of neutron star models will be  confronted with available 
observational data (Douchin \& Haensel 2001, Haensel 2001). 
}

%
%
\references{
\item
E.~Chabanat, P.~Bonche, P.~Haensel,  J.~Meyer, R.~Schaeffer. 
Nucl. Phys. A 627, 710 (1997)
\item
E.~Chabanat, P.~Bonche, P.~Haensel,  J.~Meyer, R.~Schaeffer. 
Nucl. Phys. A 635, 231  (1998)
\item
F.~Douchin, P.~Haensel, J.~Meyer. Nucl. Phys. A 665, 419 (2000)
\item
F.~Douchin, P.~Haensel. Phys. Lett. B 485, 107 (2000)
\item
F.~Douchin, P.~Haensel, J.~Meyer. Nucl. Phys. A 665, 419 (2000)
\item
P.~Haensel. Apparent radii of neutron stars, observations of Geminga and 
RX J185635-3754, and equation of state dense matter. A\&A Letters
(subm., \arxiv{astro-ph/0105485})
\item
P.~Haensel, J.L.~Zdunik, F. Douchin. Equation of state of dense matter and the 
minimum mass of cold neutron stars, in preparation (2001)
\item
C.P. Lorenz, D.G. Ravenhall, C.J.~Pethick. Phys. Rev. Lett. 70, 379 (1993)
\item
V.R. Pandharipande, D.G. Ravenhall. In: M. Soyeur et al. (Eds.), Proc. NATO 
Advanced  Research Workshop  on nuclear matter and heavy ion 
collisions, Les Houches, Pleanum, New York (1989)
}

\item[\bf 10.40--11.10]
\contribution
%
%
{Thermal relaxation in young and old neutron stars}
{\speaker{D.G.~Yakovlev}\inst{1}, O.Y.~Gnedin\inst{2}, A.Y.~Potekhin\inst{1}}
{%
\inst{1}%
Ioffe Physical Technical Institute%
\instaddr{, Politekhnicheskaya 26, St. Petersburg 194021 \\}
\inst{2}%
Institute of Astronomy%
\instaddr{, Madingley Road, Cambridge CB3 0HA, England \\}}
{yak@astro.ioffe.rssi.ru}
{http://www.ioffe.rssi.ru/astro/NSG/NSG-Pub1.html}
{http:/$\!$/www.ioffe.rssi.ru/astro/NSG/NSG-Pub1.html}
{astro-ph/0012306}{}{}

\abstext
{  
Thermal relaxation in young isolated neutron stars
and in soft X-ray transients is analyzed. 

The relaxation process in young neutron stars
lasts from 10 to 100 years (Lattimer et al.\ 1994,
Gnedin et al.\ 2001). During the relaxation
stage the effective surface temperature
of the star is fairly insensitive to physical
conditions in the stellar core, being mainly  
determined by poorly known physics of matter
of subnuclear density in the neutron star crust. 
The end of the relaxation stage  
is manifested by a drop of the surface temperature, while
subsequent thermal evolution of the star is mainly
controlled by properties of matter in the stellar
core. The temporal evolution of the surface
temperature during the relaxation epoch 
carries important information on neutrino
processes, thermal conductivity, heat capacity
and superfluidity of neutrons in the stellar crust.

Another example of the thermal relaxation is provided
by soft X-ray transients.
These sources undergo periods of intense accretion
separated by long phases of quiescence. 
Their thermal state is thought to be determined
(Brown et al.\ 1998)
by energy release (mainly due to pycnonuclear
reactions) in accreted matter (Haensel and Zdunik 1990) 
sinking in deep
layers of the stellar crust under the weight of newly
accreted material. The quasisteady thermal state is
reached (Colpi et al.\ 2001)
in about $5 \times 10^4$ yrs after the transient activity
onset. The star becomes warm, some fraction of thermal
energy flows into the core and is radiated away by
neutrinos. The effective surface temperature
in the quiescence periods becomes sensitive to
neutrino emission mechanisms and superfluidity
of matter in the stellar core, and therefore, to the
stellar mass. This gives a new method (Colpi et al.\ 2001) to measure
masses of neutron stars and superfluid transition temperatures
in stellar cores.

}

%
%
\references{

\item
Brown E.F., Bildsten L., Rutledge R.E., 1998, ApJ 504, L95
(\arxiv{astro-ph/9807179})

\item
Colpi M., Geppert U., Page D., Possenti A., 2001,
ApJ 548, L175 (\arxiv{astro-ph/0010572})

\item
Gnedin O.Y., Yakovlev D.G., Potekhin A.Y., 2001,
MNRAS, 324, 725 (\arxiv{astro-ph/0012306})

\item
Haensel P., Zdunik J.L., 1990, A\&A, 227, 431; 229, 117

\item
Lattimer J.M., Van Riper K.A., Prakash M., Prakash M., 1994,
ApJ 425, 802

}

\item[\bf 11.10--11.30]
\contribution
%
%
{Thermal structure and cooling of neutron stars}
%
%
%
{\speaker{A.Y.~Potekhin}\inst{1}, D.G.~Yakovlev\inst{1}}
{%
\inst{1}%
Ioffe Physical Technical Institute%
\instaddr{, Politekhnicheskaya 26, St. Petersburg 194021 \\}}
{palex@astro.ioffe.rssi.ru}
%
%
%
{http://www.ioffe.rssi.ru/astro/NSG/NSG-Pub1.html}
{http:/$\!$/www.ioffe.rssi.ru/astro/NSG/NSG-Pub1.html}
{astro-ph/9909100}{astro-ph/0105261}{}


\abstext
{
Thermal structure of neutron stars with magnetized envelopes
is studied using modern physics input -- in particular,
updated thermal conductivity of a dense magnetized plasma
typical for neutron-star envelopes [1].
The relation between the internal ($T_{\rm int}$)
and local surface temperatures is calculated
and fitted by analytic expressions
for magnetic field strengths $B$ from 0 to 10$^{16}$~G
and arbitrary inclination of the field lines to the surface.
The luminosity of a neutron star with dipole magnetic field
is calculated and fitted as a function
of $B$, $T_{\rm int}$, stellar mass and radius.

In addition, we simulate cooling
of neutron stars with magnetized envelopes,
using the modern cooling code [2].
In particular, we analyse magnetic field effects 
in the ultramagnetized envelopes
of magnetars on the cooling curves. 
Finally, we demonstrate that the magnetic field
of the Vela pulsar strongly affects 
observational constraints on the values of
critical temperatures of neutron and proton superfluids
in its core.
}

%
%
\references{

\item
A.Yu.~Potekhin, Astron. Astrophys., 351, 787 (1999)

\item
O.Yu.~Gnedin, D.G.~Yakovlev, A.Y.~Potekhin. 
MNRAS 2001, 324, 725 (\arxiv{astro-ph/0012306})

\item
A.Yu.~Potekhin, D.G.~Yakovlev, Astron.\& Astrophys., accepted (2001)

}

\item[\bf 11.30--11.50]
\contribution
%
%
{Masses of
isolated neutron stars and superfluid gaps \nl
in their cores: constraints from observations and \nl
cooling theory
}
%
%
%
{\speaker{A.D.~Kaminker}\inst{1}, P.~Haensel\inst{2}, D.G.~Yakovlev\inst{1}}
{%
\inst{1}%
Ioffe Physical Technical Institute,%
\instaddr{ Politekhnicheskaya 26, St. Petersburg 194021 \\}
\inst{2}%
N.~Copernicus Astronomical Center%
\instaddr{, Bartycka 18, 00-716 Warszawa, Poland\\}}
{kam@astro.ioffe.rssi.ru}
%
%
%
{http://www.ioffe.rssi.ru/astro/NSG/NSG-Pub1.html}
{http:/$\!$/www.ioffe.rssi.ru/astro/NSG/NSG-Pub1.html}
{astro-ph/0105047}{}{}


\abstext
{
We present the results of new cooling simulations
of neutron stars under the assumption
that stellar cores consist of neutrons, protons
and electrons. We assume further that
nucleons may be in a superfluid state and 
use realistic density profiles
of superfluid critical temperatures
$T_{\rm cn}(\rho)$ and $T_{\rm cp}(\rho)$ of neutrons and protons.
Taking a suitable profile of $T_{\rm cp}(\rho)$ 
with maximum $\sim 5 \times 10^9$ K we
obtain smooth transition from slow to rapid cooling
with increasing stellar mass. Adopting
the same profile we can explain the majority of observations of thermal 
emission from isolated middle--aged
neutron stars by cooling
of neutron stars with different masses  either
with no neutron superfluidity in the cores
or with a weak superfluidity, $T_{\rm cn} < 10^8$ K.
The required masses depend sensitively
on the decreasing slope of the $T_{cp}(\rho)$ profile
at $\rho \sim 10^{15}$ g cm$^{-3}$. 
For one particular model $T_{cp}(\rho)$ profile the masses range 
from $\sim 
1.2 \, {\rm M}_\odot$ for
(young and hot) RX J0822--43 and 
(old and warm) PSR 1055--52 
and RX J1856-3754
to $\approx 1.45\, {\rm M}_\odot$
for the (rather cold) Geminga and Vela pulsars.
Shifting the decreasing slope of $T_{cp}(\rho)$
profile to higher densities
we may obtain higher masses of the same sources.
This gives a new method to constrain neutron star
masses and superfluid critical temperatures in the stellar
cores.
}

%
%
\references{

\item
Kaminker A.D., Haensel P., Yakovlev D.G. 2001, A\&A Lett., in press
(\arxiv{astro-ph/0105047})

}

\item[\bf 11.50--12.10]
\contribution
%
%
{Viscous damping of neutron star pulsations}
%
%
%
{\speaker{K.P.~Levenfish}\inst{1}, P.~Haensel\inst{2}, D.G.~Yakovlev\inst{1}}
{%
\inst{1}%
Ioffe Physical Technical Institute,%
\instaddr{ Politekhnicheskaya 26, St. Petersburg 194021 \\}
\inst{2}%
N. Copernicus Astronomical Center%
\instaddr{, Bartycka 18, 00-716 Warszawa, Poland\\}}
{ksen@astro.ioffe.rssi.ru}
%
%
%
{http://www.ioffe.rssi.ru/astro/NSG/NSG-Pub1.html}
{http:/$\!$/www.ioffe.rssi.ru/astro/NSG/NSG-Pub1.html}
{astro-ph/0004183}{astro-ph/0103290}{}


\abstext
{
Bulk and shear viscosities of matter in neutron star
cores composed of nucleons, hyperons and/or quarks
are discussed taking into account the effects of
superfluidity of various particle species. 
The bulk viscosity of nonsuperfluid matter
is enhanced by 4--5 orders of magnitude
if direct Urca process of neutrino emission is open,
and it is further enhanced by several orders of
magnitude in the presence of 
hyperons and quarks. Strong superfluidity of
matter may greatly dump the bulk viscosity.
Viscous damping times of neutron star pulsations
(particularly, r-modes) are shown to be very sensitive
to composition and superfluid properties of dense matter.
In particular, this is crucial for gravitational radiation
driven instabilities.
}





\medskip
\item[\bf 12.10--12.30]
{\it Coffee break}
\bigskip

\item[\bf 11.50--12.10]
\contribution
%
%
{Accreting strange stars}
%
%
%
{\speaker{J.L. Zdunik}\inst{1}}
{%
\inst{1}%
N.~Copernicus Astronomical Center, Polish Academy of Sciences%
\instaddr{, Bartycka 18, PL-00-716 Warszawa, Poland \\}}
{jlz@camk.edu.pl}
%
%
%
{}
{}
{astro-ph/0002394}{astro-ph/0104116}{}


\abstext
{
The models of the rotating strange stars (SS) with crust
are presented in the context of the spin up by accretion.
Calculations are performed within the framework of general
relativity. Equation of state of strange quark matter is based
on the MIT Bag Model with massive strange quarks and lowest order
QCD interactions. The crust is described by the BPS
equation of state.
 
The properties of the rotating strange stars and main differences with
respect to neutron stars are discussed (the large oblateness of rotating SS,
the gravitational field in the outer spacetime, the properties of the 
innermost stable circular orbit [1,2,3]).

The evolutionary tracks for the strange stars accreting the matter from the
margi\-nally stable orbit are presented. 
If all the particle angular momentum is transferred to the star
the evolutionary tracks terminate at
mass-shed limit for almost all initial values of the stellar mass.  Only 
stars with the mass very close to the maximum one can terminate by
encountering the radial instability. 
However the collapse is the most probable fate of the
accreting SS provided that only  50\% of the particle 
angular momentum is deposited onto the star.  
The maximum amount of the accreted mass is about 0.6$\,M_{\odot}$, the 
required mass to spin up star to millisecond periods is 
$\sim$  0.2$\,M_{\odot}$.
}

%
%
\references{

\item
Zdunik J.L.,  P. Haensel, D. Rosinska, E. Gourgoulhon, 2000, A\&A 356, 612

\item
Zdunik J.L, Gourgoulhon E., 2001, Phys Rev D, 63, 087501

\item
Zdunik J.L, Haensel, P., Gourgoulhon E., 2001,  A\&A, 
in press (\arxiv{astro-ph/0104116})

}

\bigskip
\item[{\it Session 6.}]
{\it POSTERS}\\
Chairman: V.S.~Beskin

\item[1.]
\contribution
%
%
{Self-sustained neutron haloes in the inner parts \nl of hot accretion disks}
%
%
%
{\speaker{E.V.~Derishev}\inst{1}, A.A.~Belyanin\inst{1}}
{%
\inst{1}%
Institute of Applied Physics%
\instaddr{, 46 Ulyanov st., 603600, Nizhny Novgorod, Russia \\}}
{derishev@appl.sci-nnov.ru}
%
%
%
{}
{}
{}{}{}


\abstext
{
We suggest and analyze in detail the possibility of self-sustained  
neutron halo formation in the vicinity of disk-accreting black hole 
or neutron star. Initial seed neutrons originate from collisional 
dissociation of helium in a hot infalling plasma. Once the neutron halo 
is formed, it becomes ``self-sustained'', i.e.,  supported by 
collisions of energetic neutrons from the halo with helium nuclei 
even if  the  ion temperature in the disk drops below the threshold  
for He dissociation by protons. 
 
Our analysis shows that the halo formation is possible when the mass  
accretion rate is below the critical value $\dot{M}_{max} \sim$  
5$\cdot$ 10$^{16}M/M_{\odot}$~g/s, where $M$ is the black-hole mass. 
In this case neutrons accrete slower than ions which leads to the  
neutron pile-up: the density of neutrons can be much higher than the  
proton density. Under certain conditions neutrons may diffuse 
outwards into the cooler part of an accretion disk, leading to 
enhanced deuterium production accompanied by emission in a relatively 
narrow 2.2 MeV line.
 
Neutron halo exists in a broad range of disk parameters and mass 
accretion rates. We discuss the observability of various dynamical 
and radiative phenomena related to the existence of neutron halo.
}

%
%

\item[2.]
\contribution
%
%
{Magnetocavitation mechanism of gamma-ray burst \nl phenomenon}
%
%
%
{\speaker{Yu.N.~Gnedin}\inst{1}, S.O.~Kiikov\inst{1}}
{%
\inst{1}%
Central (Pulkovo) Astronomical Observatory%
\instaddr{, Pulkovskoe Shosse 65/1, 196140, \nl St.-Petersburg, Russia \\}}
{gnedin@gao.spb.ru, kiikov@gao.spb.ru}
%
%
%
{}
{}
{astro-ph/9908124}{}{}


\abstext
{We consider the phenomenon of a gamma-ray burst as a nonlinear collapse 
of a magnetic cavity surrounding a neutron star with huge magnetic 
field due to the bubble shape instability in a resonant 
MHD field of an accreting plasma or a plasma
on the neutron star surface. The QED 
effect of vacuum polarizability by a strong magnetic field is taken into 
consideration. We develop the analogy with the phenomenon of 
sonoluminescence in which the gas bubble is located in surrounding liquid 
with a driven sound intensity. 
}

%
%

\item[3.]
\contribution
%
%
{Interaction of anisotropic pulsar winds \nl 
with interstellar medium
}
%
%
%
{\speaker{D.V.~Khangoulian}\inst{1}, S.V.~Bogovalov\inst{1}}
{%
\inst{1}%
Astrophysics Institute at the Moscow state Engineering Physics Institute%
\instaddr{, Kashirskoje Shosse 31, 115409, Moscow, Russia \\}}
{bogoval@axpk40.mephi.ru}
%
%
%
{}
{}
{}{}{}


\abstext
{
The objective of the work is to explain the 
morphology of X-ray plerions around the Crab and the Vela pulsars observed 
with the Chandra telescope. The X-ray plerions consist of toroidal structure and 
jet-like features directed along the axis of rotation. We assume that the 
observed structures are produced 
due to interaction of the anisotropic relativistic winds ejected by radio 
pulsars with interstellar medium. MHD models of the wind formation predict 
that the energy flux in the wind is concentrated at the equatorial region.  
For one of the simplest distributions of the energy flux in the wind
we have calculated the shape of the terminating shock wave and 
compare it with observations.
}

%
%

\item[4.]
\contribution
%
%
{First detection of the Vela pulsar in IR with the VLT}
%
%
%
{\speaker{A.B.~Koptsevich}\inst{1}, P.~Lundqvist\inst{2},
J.~Sollerman\inst{3}, Yu.A.~Shibanov\inst{1}, S.~Wagner\inst{4}}
{%
\inst{1}%
Ioffe Physical Technical Institute,%
\instaddr{ Politekhnicheskaya 26, St. Petersburg 194021 \\}
\inst{2}%
Stockholm Observatory,%
\instaddr{ SCFAB, Stockholm Observatory, Department of Astronomy, 
SE-106 91 Stockholm, Sweden \\}
\inst{3}%
ESO,%
\instaddr{ Karl-Schwarzschild Strasse 2, D-85748, Garching bei M\"unchen,
Germany \\}
\inst{4}%
University of Heidelberg%
\instaddr{, Landessternwarte K\"onigstuhl, D-69117, Heidelberg, Germany \\}}
{kopts@astro.ioffe.rssi.ru}
%
%
%
{}
{}
{}{}{}


\abstext
{
We report detection of the Vela pulsar 
in near-infrared during observation 
performed with VLT/ISAAC on December 2000 -- January 2001.
The pulsar is clearly identified at the images in J and H bands.
Preliminary estimations show that the  emission  
is of nonthermal origin. By this detection Vela 
starts to be a forth member of a small family 
of radio pulsars (Crab, Geminga,  and PSR B0656+14)   
detected in infrared.  
}

%
%

\item[5.]
\contribution
%
%
{Familon emissivity of magnetized plasma}
%
%
%
{\speaker{E.N.~Narynskaya}\inst{1}, N.V.~Mikheev\inst{1}}
{%
\inst{1}%
Yaroslavl State University%
\instaddr{, Sovietskaya 13, 150000, Yaroslavl, Russia \\}}
{elenan@univ.uniyar.ac.ru}
%
%
%
{}
{}
{}{}{}


\abstext
{
Emission of familons in the processes $e^- \to e^-
+ f$, $ e^- \to \mu + f$ in a magnetized plasma is
investigated in the model in which familons either have 
or have not direct coupling to leptons via plasmon.
Contributions of the lowest and excited Landau levels
are analyzed. The differential probabilities and integral familon
effect on the plasma are calculated. It is shown
that in the process $e^- \to \mu + f$ P -- odd 
interference phenomenon leads to familon force
acting on  plasma along the magnetic field.
}

%
%

\item[6.]
\contribution
%
%
{Evolution of isolated neutron stars in globular clusters: \nl 
number of accretors}
%
%
%
{\speaker{S.B.~Popov}\inst{1}, M.E.~Prokhorov\inst{1}}
{%
\inst{1}%
Sternberg Astronomical Institute%
\instaddr{, Universitetskij pr. 13, 119899, Moscow, Russia \\}}
{polar@sai.msu.ru}
%
%
%
{http://xray.sai.msu.ru/~polar/}
{http:/$\!$/xray.sai.msu.ru/$\sim$polar/}
{astro-ph/0102201}{}{}


\abstext
{
With a simple model from the point of view of population 
synthesis we try to verify an
interesting suggestion made by Pfahl \& Rappaport (2000) 
that dim sources in globular clusters
(GCs) can be isolated accreting neutron stars (NSs). 
Simple estimates show, that we can expect
about $0.5-1$ accreting isolated NS per typical GC with 
$M=10^5$ M$_{\odot}$ which agrees
with observations. Properties of old accreting 
isolated NSs in GCs are briefly
discussed. We suggest that accreting NSs in GCs experienced 
significant magnetic field decay. 
}

%
%

\item[7.]
\contribution
%
%
{Some features of weak and strong pulses of pulsars}
%
%
%
{\speaker{M.Yu.~Timofeev}\inst{1}}
{%
\inst{1}%
Astro Space Centre FIAN%
\instaddr{, 117810, Moscow, Russia \\}}
{timofeev@anubis.asc.rssi.ru}
%
%
%
{}
{}
{}{}{}


\abstext
{
This presentation is based on the observations, which have been 
made in Effelsberg with the 100-meter radio telescope
in April 1994. Recording was produced in the mode of
the single pulses registration of five pulsars: PSR
0823+26, PSR 0950+08, PSR 1706$-$16, PSR 1929+10, PSR
2021+51 at 1.7 GHz with different time resolutions.

The goal of the programme is the investigation of some
features of individual pulses. Parameter $q$ was calculated
for any single pulse. To do this, the maximum of
intensity inside of pulse ($I_m$) was found for any
pulse and it was normalized by r.m.s. calculated
outside the pulse. Single radio pulses were divided into
groups according to the values of $q$, and thus average
profiles of total intensity in the chosen region were calculated.

It is discovered that the width of an average profile of
total intensity emission decreases while $q$ increases.
Also we have obtained some time shifts between average profiles
with different $q$. Then Gaussian fits were made for
any pulsar profile and the time shift and width
parameters were refined.

The explanation of these effects can be the following. 
Pulsar radiation is generated in a 
magnetospheric layer. Strong pulses are generated at the inner
side of the layer, close to the pulsar surface, 
where generation of the plasma oscillations may be expected. 
As the height increases, the
synchronization condition is violated 
which
leads to the emission of weaker pulses.
}

%
%

\medskip
\item[\bf 13.40--14.40]
{\it Lunch}
\bigskip


\item[{\it Session 7.}]
{\it OBSERVATIONS OF ISOLATED NEUTRON STARS}\\
Chairman: B.V.~Komberg

\item[\bf 14.40--15.10]
\contribution
%
%
{Multiwavelength observations of isolated NSs: \nl
thermal emission vs nonthermal}
%
%
%
{\speaker{Yu.A.~Shibanov}\inst{1}}
{%
\inst{1}%
Ioffe Physical Technical Institute%
\instaddr{, Politekhnicheskaya 26, St. Petersburg 194021 \\}}
{shib@stella.ioffe.rssi.ru}
%
%
%
{}
{}
{}{}{}


\abstext
{
Multiwavelength observations are important tool  
to study the properties of thermal emission from 
cooling surfaces of isolated neutron stars (NSs), to 
distinguish thermal emission from nonthermal emission
of their magnetospheres, to understand the    
mechanisms of the magnetospheric radiation 
in different wave bands, and to investigate structure 
and properties of pulsar nebulae forming due to interaction of NSs 
with ambient matter. A wealth of new information on the multiwavelength 
radiation from radio pulsars as well as from radio quiet isolated NSs 
has been obtained in recent years year with the Chandra and XMM-Newton 
X-ray observatories, HST, and new generation of large ground 
based optical telescopes VLT and Subaru.  
We review recent results obtained in a wide 
frequency range, from IR, through optical, X-ray, to gamma-ray bands, 
discuss their implications, and prospects of further studies 
of isolated NSs.    
}

%
%

\item[\bf 15.10--15.30]
\contribution
%
%
{Multiband photometry of the PSR B0656+14 \nl and its neighborhood}
%
%
%
{\speaker{A.B.~Koptsevich}\inst{1}, G.G.~Pavlov\inst{2},
S.V.~Zharikov\inst{3}, V.V.~Sokolov\inst{4}, Yu.A.~Shibanov\inst{1}, V.G.~Kurt\inst{5}}
{%
\inst{1}%
Ioffe Physical Technical Institute,%
\instaddr{ Politekhnicheskaya 26, St. Petersburg 194021 \\}
\inst{2}%
The Pennsylvania State University,%
\instaddr{ Department of Astronomy \& Astrophysics, 525 Davey Lab,
University Park, PA 16802, USA; \\}
\inst{3}%
Obs. Astr. Nacional. de Inst. de Astronomia de UNAM,%
\instaddr{ Observatorio Astronomico Nacional de Instituto de Astronomia de UNAM,
Ensenada,  B.C., 228, Mexico \\}
\inst{4}%
SAO RAS,%
\instaddr{ Nizhny Arkhyz, 357147, Karachai-Cherkessia, Russia \\}
\inst{5}%
Astro Space Centre FIAN%
\instaddr{, 117810, Moscow, Russia \\}}
{kopts@astro.ioffe.rssi.ru}
%
%
%
{http://www.ioffe.rssi.ru/astro/NSG/obs/0656-phot.html}
{http:/$\!$/www.ioffe.rssi.ru/astro/NSG/obs/0656-phot.html}
{astro-ph/0009064}{}{}


\abstext
{
We present the results of broad-band photometry
of the nearby middle-aged radio pulsar PSR B0656+14 and its neighborhood
obtained with the 6-meter telescope of the SAO RAS and with the {\sl
Hubble Space Telescope}.
The broad-band
spectral flux $F_\nu$
of the pulsar decreases with increasing frequency in the near-IR
range and increases with frequency in the near-UV range.
The increase towards UV can
be naturally interpreted as
the Rayleigh-Jeans tail of the soft thermal
component of the X-ray spectrum emitted
from the
surface of the cooling neutron star.
Continuation of the power-law component, which dominates
in the
high-energy tail of the
X-ray spectrum, to the IR-optical-UV frequencies is consistent with the
observed fluxes.
This suggests that
the non-thermal pulsar radiation may be of the same origin
in a broad frequency range from IR to hard X-rays.
We also studied 4 objects detected within
5" from the pulsar.
}

\references{
\item
Koptsevich A.B., Pavlov G.G., Zharikov S.V., Sokolov V.V., Shibanov Yu.A., Kurt V.G.
2001, A\&A, 370, 1004
}

\item[\bf 15.30--15.50]
\contribution
%
%
{Multiband photometry of Geminga}
%
%
%
{\speaker{V.N.~Komarova}\inst{1}, V.G.~Kurt\inst{2}, T.A.~Fatkhullin\inst{1},
V.V.~Sokolov\inst{1}, Yu.A.~Shibanov\inst{3}, A.B.~Koptsevich\inst{3}}
{%
\inst{1}%
SAO RAS,%
\instaddr{ Nizhny Arkhyz, 357147, Karachai-Cherkessia, Russia  \\}
\inst{2}%
Astro Space Centre FIAN,%
\instaddr{ 117810, Moscow, Russia \\}
\inst{3}%
Ioffe Physical Technical Institute%
\instaddr{, Politekhnicheskaya 26, St. Petersburg 194021 \\}}
{vkom@sao.ru}
%
%
%
{}
{}
{}{}{}


\abstext
{
The results of Geminga's investigation with the aid of 
BVRI-photometry based on observations with the 6\,m 
telescope are presented. The obtained Geminga's      
magnitudes B (26.1$\pm$0.5), V (25.3$\pm$0.3) and 
R$_c$ (25.4$\pm$0.3) are consistent with the results of 
other studies of its optical emission in this spectral 
range. We derive for the first time the magnitude in 
I$_c$-band to be $25^m.1\pm0.4$, which is more than 
a magnitude higher if compared with the upper limit 
given in the paper by Bignami et al. (1996).                   
The comparison of the broadband spectra of this 
middle-aged isolated neutron star with the results of 
observations in X-rays and hard ultraviolet allows us
to assume the presence of the features probably of nonthermal 
character in Geminga's emission at least in some 
regions of the visible spectrum.                    
}

%
%

\item[\bf 15.50--16.10]
\contribution
%
%
{ Separation of millisecond and submillisecond 
pulsations \nl in hard gamma radiation 
from the Geminga pulsar}
%
%
%
{\speaker{E.I.~Chuikin}\inst{1}}
{%
\inst{1}%
Ioffe Physical Technical Institute%
\instaddr{, Politekhnicheskaya 26, St. Petersburg 194021 \\}}
{evgeniy.chuikin@pop.ioffe.rssi.ru }
%
%
%
{}
{}
{}{}{}


\abstext
{
Separation of millisecond and submillisecond pulsations in radiation from 
the Geminga pulsar
(PSR J0633+1746) 
with ``Discrete temporal Fourier 
series analysis method'' on the data base of GAMMA-1 observations (from 30 
November 1990 to 2 February 1991) is discussed. The pulsations are analysed 
also with a "Superposition of epoch method".
Significant pulsations were found within the first peak of the Geminga
light curve. Among the strongest ones there 
are the pulsations with period values
6.1313 ms, 1.7060 ms, 0.3966 ms. 
As in the previous studies of pulsations of the Vela pulsar,
reliability of the results obtained for Geminga was confirmed by 
the phasogram pattern and phase of the peak being stable
throughout the observational set.
}

%
%
%
%

\medskip
\item[\bf 16.10-16.40]
{\it Coffee break}
\bigskip

\item[{\it Session 8.}]
{\it DISCS, QPOs, MAGNETOSPHERES}\\
Chairman: K.A.~Postnov

\item[\bf 16.40--17.10]
\contribution
%
%
{Spectral and timing  properties of neutron star and \nl
black hole systems.  Theory and observations}
%
%
%
{\speaker{L.G.~Titarchuk}\inst{1}}
{%
\inst{1}%
George Mason University and Naval Research Laboratory%
\instaddr{, Center for Earth Observing and Space
          Research, George Mason University, Fairfax, VA, 22030; \\
          US Naval Research Laboratory, Washington, DC 20375-5352  \\}}
{lev@xip.nrl.navy.mil}
%
%
%
{}
{}
{}{}{}


\abstext
{
I  present  continued development of a new theoretical 
paradigm,  the Transition Layer    Model (TLM),  
for explaining features of power density 
spectra (PDS) in accreting neutron stars and black holes.   
The prevailing model for all accreting  neutron stars has been 
that an ionized  disk extends inward in the equatorial plane  
until it encounters the 
magnetosphere at the Alfven radius, inside which  MHD effects 
govern the structure and flow.  This paradigm was developed 
originally for binary pulsars. An attempt was made to extend it 
to low-mass binary systems (LMXBs) started after discovery of QPOs by EXOSAT. 
 The attempted extension  has encountered many difficulties. In the TLM 
paradigm the disk is only  partially conducting 
and its structure inside a critical radius,  
R$_{outer}$,  is governed by fluid physics.   
In TLM a pair of fluid oscillatory  modes that 
become key features related to the observed PDS.  
The  TLM has been developed through scrutiny 
of the full range of observational 
knowledge available concerning both the energy spectra and 
the PDS of neutron star and black hole sources.  
Successes include detailed fits to  the variation of 6 PDS 
features in four  LMXB sources with at most one free parameter, 
and even that one constrained. 
It is important to understand the limits of success of such a  
radical paradigm shift.  This talk covers continued development 
of  frontier areas of the theory as well as continuation of the 
crucial detailed  confrontation with observational data, done 
within the TLM  paradigm.  
Under this study both LMXB and black hole sources are studied.   
(It is possible that the paradigm will eventually extend even to the binary 
pulsars that were the starting point of the older paradigm).    
}

%
%

\item[\bf 17.10--17.30]
\contribution
%
%
{The origin of QPOs of X-ray binaries}
%
%
%
{\speaker{Yu.N.~Gnedin}\inst{1}, S.O.~Kiikov\inst{1}}
{%
\inst{1}%
Central (Pulkovo) Astronomical Observatory%
\instaddr{, Pulkovskoe Shosse 65/1, 196140, \nl St.-Petersburg, Russia \\}}
{gnedin@gao.spb.ru, kiikov@gao.spb.ru}
%
%
%
{}
{}
{}{}{}


\abstext
{We present three mechanisms of generation of X-ray quasi-periodic 
oscillations (QPOs) in binaries. Two of them are based on the analogy 
to nonlinear oscillations of gaseous caverns in liquid. The first 
mechanism is called magnetocavitation;
it implies that X-ray QPOs of a neutron 
star are produced by radial oscillations of the neutron star magnetosphere 
interacting with accreting plasma. To study X-ray QPOs of neutron stars 
with critical (Eddington) luminosity, the photon-cavitation mechanism has 
been considered. In this case X-ray QPOs of the neutron star are generated 
by radial oscillations of photon caverns in a fully ionized hydrogen 
plasma settling in the accreting column of the compact object. To explain 
X-ray QPOs of binaries with a black hole and cataclysmic variables we suggest 
the mechanism according to which X-ray QPOs are caused by nonlinear 
oscillations of current sheets originating in accreting disks. The calculated 
values of basic physical parameters of QPOs such as basic frequencies, 
frequency and amplitude dependence on the X-ray flux level and the energy of 
photons and also on QPOs lag time of photons of different energies are 
consistent with the observational data.
}

%
%

\item[\bf 17.30--17.50]
\contribution
%
%
{On the 2D structure of a thin accretion disk}
%
%
%
{\speaker{V.S.~Beskin}\inst{1}, R.Yu.~Kompaneetz\inst{2}}
{%
\inst{1}%
Lebedev Physical Institute,%
\instaddr{ 142292, Moscow, Russia \\}
\inst{2}%
Moscow Institute of Physics and Technology%
\instaddr{, Dolgoprudnyi 141700, Russia \\}}
{beskin@lpi.ru}
%
%
%
{}
{}
{}{}{}


\abstext
{
The internal structure of a thin accretion disk 
in the vicinity of a black hole inside the last stable orbit 
($R<$ 3$R_g$) is considered within the hydrodynamical version of 
the Grad-Shafranov equation. It is shown that at any way up to 
the sonic surface the flow is not radial, the flow parameters 
in the equatorial region (e.g., the poloidal velocity) differing 
from those near the disk boundary. 
}

\item[\bf 17.50--18.10]
\contribution
%
%
{Gamma rays from pulsar winds}
%
%
%
{\speaker{S.V.~Bogovalov}\inst{1}, F.A.~Aharonian\inst{2}}
{%
\inst{1}%
Astrophysics Institute at the Moscow state Engineering Physics Institute,%
\instaddr{ Kashirskoje Shosse 31, 115409, Moscow, Russia \\}
\inst{2}%
Max-Planck-Institut f\"ur Kernphysik%
\instaddr{, Postfach 103980, D-69029, Heidelberg, Germany  \\}}
{bogoval@axpk40.mephi.ru, Felix.Aharonian@mpi-hd.mpg.de}
%
%
%
{}
{}
{astro-ph/0003157}{}{}


\abstext
{
The relativistic winds of  pulsars with average Lorentz-factor
$\gamma_W \sim$ 10$^6$ could be directly observed through its inverse Compton
(IC) $\gamma$-ray emission caused by illumination of the wind by
low-frequency photons emitted by the surface of the pulsar and its
magnetosphere.  Calculations performed for the Crab and Vela pulsars
show that even the interaction of the wind electrons with thermal emission of
the neutron star gives detectable fluxes of the gamma-ray emission.
Comparison of the observations of the Crab pulsar in gamma-rays above
0.5 TeV with calculations gives rather
important constraints on the mechanism of the wind acceleration.
To avoid overproduction of the pulsed TeV emission from Crab, the wind
should be accelerated at a distance exceeding 30 light cylinders in
conventional models of radio pulsars. The acceleration of the wind
close to the light cylinder
is possible only in the presence of
rather strong constraints on the mechanisms of 
generation of soft nonthermal radiation by the pulsar.
}

%
%

\item[\bf 18.10--18.30]
\contribution
%
%
{Interaction of anisotropic pulsar winds \nl 
with interstellar medium
}
%
%
%
{\speaker{D.V.~Khangoulian}\inst{1}, S.V.~Bogovalov\inst{1}}
{%
\inst{1}%
Astrophysics Institute at the Moscow state Engineering Physics Institute%
\instaddr{, Kashirskoje Shosse 31, 115409, Moscow, Russia \\}}
{bogoval@axpk40.mephi.ru}
%
%
%
{}
{}
{}{}{}


\abstext
{
The objective of the work is to explain the 
morphology of X-ray plerions around the Crab and the Vela pulsars observed 
with the Chandra telescope. The X-ray plerions consist of toroidal structure and 
jet-like features directed along the axis of rotation. We assume that the 
observed structures are produced 
due to interaction of the anisotropic relativistic winds ejected by radio 
pulsars with interstellar medium. MHD models of the wind formation predict 
that the energy flux in the wind is concentrated at the equatorial region.  
For one of the simplest distributions of the energy flux in the wind
we have calculated the shape of the terminating shock wave and 
compare it with observations.
}

%
%

\bigskip
\begin{center}
\large\bf June 8, Friday
\end{center}
\medskip

\item[{\it Session 9.}]
{\it NEUTRON STARS IN THE GALAXY AND BEYOND}\\
Chairman: Yu.N.~Gnedin

\item[\bf 10.00--10.40]
\contribution
{Wolf-Rayet stars and relativistic objects}
{\speaker{A.M.~Cherepashchuk}\inst{1}}
{%
\inst{1}%
Sternberg Astronomical Institute%
\instaddr{, Universitetskij pr. 13, 119899, Moscow, Russia \\}} 
{cher@sai.msu.ru} 
{} 
{}
{astro-ph/0012512}{}{}

\abstext{
The observed properties of Wolf-Rayet stars and
relativistic objects in close binary systems are analyzed. The
final masses $M^{f}_{\rm CO}$ for the carbon-oxygen cores of WR
stars in WR+O binaries are calculated taking into account the
radial loss of matter via stellar wind, which depends on the mass
of the star. The analysis includes new data on the clumpy
structure of WR winds, which appreciably decreases the required
mass-loss rates $M_{\rm WR}$ for the WR stars. The masses
$M^{f}_{\rm CO}$ lie in the range (1--2)$M_\odot$ -- (20--44)$M_\odot$
and have a continuous distribution. The masses $M_x$ of the
relativistic objects are (1--20)$M_\odot$ and have a bimodal
distribution: the mean masses for neutron stars and black holes
are (1.35 $\pm $0.15$)M_\odot$ and (8--10)$M_\odot$, respectively, with
a gap from (2--4)$M_\odot$ in which no neutron stars or black holes
are observed in close binaries. The mean final CO-core mass is
$\overline M^f_{\rm CO}$ = (7.4--10.3)$M_{\odot}$, close to the mean
mass for the black holes. This suggests that it is not only the
mass of the progenitor that determines the nature of the
relativistic object, but other parameters as well --- rotation,
magnetic field, etc. One SB1R Wolf-Rayet binary and 11 suspected
WR+C binaries that may have low-mass X-ray binaries with neutron
stars and black holes.
}

\item[\bf 10.40--11.00]
\contribution
%
%
{Distribution of NS and BH masses \nl and mechanism of 
SN explosion}
%
%
%
{\speaker{M.E.~Prokhorov}\inst{1}, K.A.~Postnov\inst{1}}
{%
\inst{1}%
Sternberg Astronomical Institute%
\instaddr{, Universitetskij pr. 13, 119899, Moscow, Russia \\}}
{mike@sai.msu.ru}
%
%
%
{}
{}
{}{}{}


\abstext
{
The observed distribution of masses of compact remnants of
 massive star evolution (neutron stars and black holes)
 is analysed and its relation with plausible core collapse
 supernova mechanisms is discussed. It is argued that
 the observed absence of compact stars with masses 1.5--3.0 $M_\odot$
 is in favor of the magnetorotational mechanism of supernova explosion
 and soft equation of state of neutron star matter with a
 limiting mass near 1.5 $M_\odot$. Observational consequences of
 this hypothesis are discussed.
}

%
%

\item[\bf 11.00--11.20]
\contribution
%
%
{Isolated neutron stars as X-ray sources: \nl
accretion vs. cooling}
%
%
%
{\speaker{S.B.~Popov}\inst{1}, M.E.~Prokhorov\inst{1},
M.~Colpi\inst{2}, R.~Turolla\inst{3}, A.~Treves\inst{4}}
{%
\inst{1}%
Sternberg Astronomical Institute,%
\instaddr{ Universitetskij pr. 13, 119899, Moscow, Russia \\} 
\inst{2}%
University Milano-Bicocca,%
\instaddr{ Paizza della Scienza 3, 20126, Milano, Italy \\}
\inst{3}%
University of Padova,%
\instaddr{ Via Marzolo 8, 35131, Padova, Italy \\}
\inst{4}%
University of Como%
\instaddr{, Via Lucini 3, 22100, Como, Italy\\}}
{polar@sai.msu.ru}
%
%
%
{http://xray.sai.msu.ru/~polar/}
{http:/$\!$/xray.sai.msu.ru/$\sim$polar/}
{astro-ph/0009225}{astro-ph/0011564}{astro-ph/0101031}


\abstext
{
We briefly review observational appearance of
isolated neutron stars which are not observed as
normal radio pulsars.

In some details we discuss dim X-ray sources
in globular clusters. We present a simple 
population synthesis model 
of old isolated neutron stars in
globular clusters to test suggestion made by Pfahl 
and Rappaport (2000), that these objects are powered
by accretion of interstellar medium onto isolated 
neutron stars.

We discuss different interpretations (cooling vs. 
accretion) of 7 dim ROSAT sources 
("The Magnificent Seven"). Log N- Log S distribution
of accreting and cooling neutron stars is obtained.
We suggest, that most of these sources are young
cooling neutron stars. Otherwise magnetic field decay
is necessary.

Influence of investigation of neutron stars in
dim X-ray sources on nearby fields of astrophysics
is briefly mentioned.
}

%
%

\item[\bf 11.20--11.40]
\contribution
%
%
{The manifestations of different acceleration mechanisms \nl
of binaries containing neutron stars}
%
%
%
{\speaker{A.I.~Tsygan}\inst{1}, V.D.~Palshin\inst{1}}
{%
\inst{1}%
Ioffe Physical Technical Institute%
\instaddr{, Politekhnicheskaya 26, St. Petersburg 194021 \\}}
{tsygan@astro.ioffe.rssi.ru, val@pop.ioffe.rssi.ru}
%
%
%
{}
{}
{}{}{}


\abstext
{
We present the results of calculations of the distribution function
of low-mass binary systems containing a neutron star
over the orbital angular momentum orientation
for different heights $z$ above the Galactic plane.
We consider two acceleration mechanisms of the binaries:
1) acceleration by their own X-ray radiation
at the stage of intense accretion of matter onto a neutron star
with asymmetric magnetic field~[1],
2) acceleration due to a supernova explosion.

In the first case the binaries are accelerated by asymmetric X-ray radiation.
The radiative reaction force $F=\xi L_X/c$ depends on
X-ray luminosity of the neutron star, $L_X$, and on the X-ray asymmetry 
parameter $\xi$. The force points along the spin axis.
For the disk accretion onto the star, its rotational axis
aligns quickly perpendicular to the accretion disk (and the orbital plane),
and the radiative reaction force will accelerate the binary along its orbital
angular momentum. 
Thus, the systems with orbital angular momentum parallel to
the Galactic axis will tend to be at large Galactic heights
while the systems with orbital angular momentum perpendicular
the Galactic axis will be accumulated in the Galactic plane.
This is in contrast to the case of symmetric supernova explosion.
For asymmetric one, the orbital angular momentum 
orientation depends mainly on the ratio of kick velocity to 
orbital velocity of neutron star.

We use the following system parameters: 
$m_1=3M_\odot$, the pre-supernova mass;
$m_2=1M_\odot$, the companion mass; 
$m_3=1.4M_\odot$, the neutron star mass;
$P=1$~day, the orbital period 
(the relative orbital velocity $v=290$~km~s$^{-1}$).
The distribution of the kick velocities ${\bf v}_k$ 
is assumed to be Maxwellian:
$(2\pi \kappa^2)^{-3/2} \exp [-u_k^2/2\kappa^2]$, 
where ${\bf u}_k\equiv \bf{v}_k/v$ is
the dimensionless kick velocity, and $\kappa$ is the dispersion of $u_k$.

To obtain the distribution functions for different $z$
we have numerically solved the equations of motion of binaries
in the Galactic potential.
At the initial moment of time the systems 
are assumed to be in the Galactic plane,
on the orbit of the Sun.

The computations show that in case of radiative 
acceleration (for $\xi=0.1,\,0.2$)
there is a linear increase of cosine of angle between the orbital angular
momentum and the Galactic axis with increasing $z$.
In case of asymmetric supernova explosion (for $\kappa=0.2$)
the distributions of the survived binary systems over 
directions of orbital angular
momentum (for different $z$) is nearly isotropic.

The results may be valid for binary millisecond radiopulsars
if they passed the stage of intense accretion.
The possibility of pulsars recycling (to produce the millisecond pulsars)
due to disk accretion in binary systems was first considered in~[2].
}

%
%
\references{

\item
V.D.~Pal'shin, A.I.~Tsygan. Astronomy Letters 24, 131 (1998)

\item
G.S.~Bisnovatyi-Kogan, B.V.~Komberg. Pis'ma v Astron Zhurn. 2, 338 (1976)
}

\item[\bf 11.40--12.00]
\contribution
%
%
{The evolution of binary stars in a globular cluster}
%
%
%
{\speaker{A.G.~Kuranov}\inst{1}, K.A.~Postnov\inst{2},
M.E.~Prokhorov\inst{2}}
{%
\inst{1}%
Faculty of Physics, Moscow State University,%
\instaddr{ 119899 Moscow, Russia \\}
\inst{2}%
Sternberg Astronomical Institute%
\instaddr{, Universitetskij pr. 13, 119899, Moscow, Russia \\}}
{alex@xray.sai.msu.ru}
%
%
%
{}
{}
{}{}{}


\abstext
{
We  present a model for the 
evolution of binary populations in a globular
cluster. The evolution of a population of close binaries is traced in a
collisional environment of evolving single stars. We consider primordial
binaries, formed simultaneously with the single stars, as well as tidally
captured binaries formed from encounters between stars. Any binary evolves
due to internal processes (e.g. evolution of its components, stellar winds,
mass transfer, etc.) and due to encounters with single stars of the
clusters. We trace individual histories of all binaries under the action 
of different physical processes, such as mass segregation,
scattering recoil, escape from the cluster. The results of our calculations
can be applied to study formation and evolution of LMXBs and millisecond
pulsars in globulars.        	
}

%
%

\medskip
\item[\bf 12.00--12.30]
{\it Coffee break}
\bigskip

\item[{\it Session 10.}]
{\it NEUTRINO PROCESSES}\\
Chairman: G.S.~Bisnovatyi-Kogan

\item[\bf 12.30--13.00]
\contribution
%
%
{Efficiency of $e^+e^-$ pair production by neutrinos} 
%
%
%
{\speaker{A.A.~Gvozdev}\inst{1}, I.S.~Ognev\inst{1}}
{%
\inst{1}%
Yaroslavl State University%
\instaddr{, Sovietskaya 13, 150000, Yaroslavl, Russia \\}}
{gvozdev@univ.uniyar.ac.ru} 
%
%
%
{}
{}
{}{}{}


\abstext
{
The dominating processes of the neutrino production and
$e^+ e^-$-pair creation by such neutrinos are 
investigated in the model of hyper-accreting disk around 
a Kerr black hole. These processes are studied both 
in vacuum  and  in the presence of strong magnetic 
field. It is shown that the efficiency of $e^+ e^-$-plasma 
production is too small to explain the 
origin of cosmological GRBs.
}

%
%

\item[\bf 13.00-13.20]
\contribution
{Three-vertex loop processes in strong magnetic field}
{\speaker{A.V.~Kuznetsov}\inst{1}, M.V.~Chistyakov\inst{1}, N.V.~Mikheev\inst{1}}
{%
\inst{1}%
Division of Theoretical Physics, Yaroslavl State University%
\instaddr{, Sovietskaya 13, 150000, \nl Yaroslavl, Russia \\}}
{avkuzn@uniyar.ac.ru}
{}
{}
{hep-ph/9808246}{hep-ph/9804444}{}

\abstext
{
A general analysis of the three-vertex loop amplitude in
very strong magnetic field, based on the asymptotic form of 
the electron propagator in the strong field, is performed.
In order to investigate the photon-neutrino process
$\gamma \gamma \to \nu \bar\nu $ and the photon splitting
$\gamma \to \gamma \gamma$, the vertex combinations of the
scalar--vector--vector ($SVV$), pseudoscalar--vector--vector ($PVV$),
3--vector ($VVV$), and axial-vector--vector--vector ($AVV$)
types are considered. It is shown that only the $SVV$ amplitude grows
linearly with the magnetic field strength, while in the
other amplitudes, $PVV$, $VVV$, and $AVV$, the linearly growing
terms are exactly canceled.
The process $\gamma \gamma \to \nu \bar\nu$ is investigated 
in the left-right-symmetric extension of the standard model 
of electroweak interaction, where the effective scalar $\nu \nu e e$
coupling is possible. 
Using the $VVV$ amplitude, the process of the photon
splitting $\gamma \to \gamma \gamma$ is investigated both below and above
the pair creation threshold.
The splitting probability is calculated taking account of the photon
dispersion and large radiative corrections near the resonance.
Possible astrophysical manifestations of the considered
processes are discussed.
Some previous results were published in the references below.
}

\references{

\item
Phys. Lett. B {\bf 434}, 67 (1998) (\arxiv{hep-ph/9804444}). 

\item
Yad. Fiz. {\bf 62}, 1638 (1999)
[Phys. At. Nucl. {\bf 62}, 1535 (1999)].

\item
In: {\it Proc. Ringberg Euroconf. "New Trends in Neutrino Physics"},
Ed. by B. Kniehl et al., 
World Sci. (1999), p. 245 (\arxiv{hep-ph/9808246}).

}

\item[\bf 13.20--13.40]
\contribution
%
%
{Photon-neu\-trino processes in a strongly \nl
magnetized plasma}
%
%
%
{\speaker{M.V.~Chistyakov}\inst{1}, N.V.~Mikheev\inst{1}}
{%
\inst{1}%
Yaroslavl State University%
\instaddr{, Sovietskaya 13, 150000, Yaroslavl, Russia \\}}
{mch@univ.uniyar.ac.ru}
%
%
%
{}
{}
{hep-ph/9907345}{}{}


\abstext
{
The effect of strongly magnetized electron-positron plasma on
photon-neutrino processes $\nu \to \nu \gamma$ and $\gamma \nu
\to \nu $ is investigated. The amplitudes and 
probabilities of the processes are calculated taking into account 
the photon dispersion and large radiative corrections near the
resonance. It is shown that the combined effect of plasma and
strong magnetic field decreases the probability of the process
$\nu \to \nu \gamma$  in comparison with the value obtained 
neglecting the plasma effect.
The probability of the process $\gamma \nu \to \nu$ does not
depend on the initial neutrino energy and is negligibly small in
comparison with probability $W_{\nu \to \nu \gamma}$ in the low
temperature limit.
}

%
%

\item[\bf 13.40--14.00]
\contribution
%
%
{Lepton pair production by high-energy neutrino \nl
in an external electromagnetic field}
%
%
%
{\speaker{D.A.~Rumyantsev}\inst{1}, A.V.~Kuznetsov\inst{1}, N.V.~Mikheev\inst{1} }
{%
\inst{1}%
Yaroslavl State University%
\instaddr{, Sovietskaya 13, 150000, Yaroslavl, Russia \\}}
{rda@uniyar.ac.ru}
%
%
%
{}
{}
{hep-ph/0003216}{}{}


\abstext
{
The process of the lepton pair production ($e^- e^+, 
e^- \mu^+, \dots$) by a neutrino 
propagating in an external electromagnetic field is 
investigated in the framework of the standard model. 
Relatively simple exact expression for the probability 
of the process
as the one-dimensional integral is obtained; it is suitable 
for a quantitive analysis.
}

%
%

\medskip
\item[\bf 14.00--15.00]
{\it Lunch}
\bigskip


\item[{\it Session 11.}]
{\it GAMMA-RAY BURSTS. Part 1}\\
Chairman: A.M.~Cherepashchuk

\item[\bf 15.00--15.30]
\contribution
%
%
{Relation of gamma-ray bursts to formation of compact stars}
%
%
%
{\speaker{K.A.~Postnov}\inst{1}}
{%
\inst{1}%
Sternberg Astronomical Institute%
\instaddr{, Universitetskij pr. 13, 119899, Moscow, Russia \\}}
{pk@sai.msu.ru}
%
%
%
{}
{}
{}{}{}


\abstext
{
We review current observational evidence
and theoretical considerations on 
formation of relativistic compacts
stars during cosmic gamma-ray bursts.
}

%
%

\item[\bf 15.30--15.50]
\contribution
%
%
{GRB as the result of interaction of supernova ejecta \nl
with NS companion of a closed binary
}
%
%
%
{\speaker{B.V.~Komberg}\inst{1}, Ya.N.~Istomin\inst{2}}
{%
\inst{1}%
Astro Space Centre FIAN,%
\instaddr{ 117810, Moscow, Russia \\}
\inst{2}%
Lebedev Physical Institute%
\instaddr{, 142292, Moscow, Russia \\}}
{bkomberg@ASC.rssi.ru}
%
%
%
{}
{}
{}{}{}


\abstext
{
The supernova explosion (I b/c or I bw  type) in a closed binary
(with separation of 10$^{13}$ cm) can produce
magnetospheric flare possessing the properties of a GRB. According
to our estimates, the
NS magnetosphere intercepts 10$^{-11}$ of the full kinetic energy
of the blast wave (10$^{47}$ ergs). Extended magnetospheric tail 
(10$^{10}$ cm)
with the mean magnetic field 10$^6$ Gauss is the source of 
gamma-rays ejected into a
small solid angle (0.1 rad). This radiation has the synchrotron nature 
being produced by accelerated particles ($E\sim1$ Gev). Fast electrons appear
as a result of the magnetic field reconnection in the current layer,
directed along the shock velocity. Such a model explains naturally 
high anisotropy of gamma-ray emission and connection of GRBs with
phenomena of compact SNe.
}

%
%

\item[\bf 15.50--16.10]
\contribution
%
%
{Short gamma ray bursts}
%
%
%
{\speaker{D.D.~Frederiks}\inst{1}}
{%
\inst{1}%
Ioffe Physical Technical Institute%
\instaddr{, Politekhnicheskaya 26, St. Petersburg 194021 \\}}
{fred@pop.ioffe.rssi.ru }
%
%
%
{}
{}
{}{}{}


\abstext
{
A review of ``short GRB'' observations with Konus instruments is given.
An observational base of this intriguing class of bursts contains now
more than 100 events. General properties of short GRBs are considered in
comparison with long events. Some interesting bursts are discussed
in detail.
}

%
%

\item[\bf 16.10--16.30]
\contribution
%
%
{On the competition between synchrotron and Compton \nl
emission of electrons in relativistic shocks}
%
%
%
{\speaker{E.V.~Derishev}\inst{1}, V.V.~Kocharovsky\inst{1}, Vl.V.~Kocharovsky\inst{1}}
{%
\inst{1}%
Institute of Applied Physics%
\instaddr{, 46 Ulyanov st., 603600, Nizhny Novgorod, Russia \\}}
{derishev@appl.sci-nnov.ru}
%
%
%
{}
{}
{}{}{}


\abstext
{
We develop a self-consistent theory of the 
synchrotron-self-Compton 
emission for optically thin sources with a stationary injection of 
relativistic mono-energetic electrons. We investigate the electron 
distribution function, synchrotron and inverse Compton (IC) spectra 
and find their analytical asymptotes. It is shown that
self-consistent steady-state 
electron distribution may produce a variety of low-energy spectral 
indices ranging from 1/2 to 1. 

The spectrum of comptonized radiation is entangled with low-energy 
synchrotron emission via the electron distribution. We derive explicit 
expression allowing one to find IC spectrum by means of the integral 
transformation of synchrotron spectrum.

Applying this theory to Gamma-Ray Burst sources we find new 
arguments in favor of synchrotron origin of the observed sub-MeV 
radiation.
}

%
%

\item[\bf 16.30--16.50]
\contribution
%
%
{Gravitational wave background from coalescing compact \nl
stars in eccentric orbits}
%
%
%
{\speaker{V.B.~Ignatiev}\inst{1}, A.G.~Kuranov\inst{2}, K.A.~Postnov\inst{2},
M.E.~Prokhorov\inst{2}}
{%
\inst{1}%
Faculty of Physics, Moscow State University,%
\instaddr{ 119899 Moscow, Russia \\}
\inst{2}%
Sternberg Astronomical Institute%
\instaddr{, Universitetskij pr. 13, 119899, Moscow, Russia \\}}
{whirl@xray.sai.msu.ru}
%
%
%
{}
{}
{astro-ph/0106299}{}{}


\abstext
{
Stochastic gravitational wave background produced by a 
stationary coalescing population of binary neutron 
stars in the Galaxy is calculated. This background is 
found to constitute a confusion limit within the LISA 
frequency band up to a limiting frequency 
$\nu_{lim}\sim $10$^{-3}$~Hz, leaving the frequency 
window $\sim $10$^{-3}$--10$^{-2}$~Hz open for
potential detection of cosmological stochastic 
gravitational waves.
}

\medskip
\item[\bf 16.50-17.20]
{\it Coffee break}
\bigskip

\item[{\it Session 12.}]
{\it GAMMA-RAY BURSTS. Part 2}\\
Chairman: Yu.A.~Shibanov

\item[\bf 17.20--17.50]
\contribution
%
%
{On the gamma-ray burst progenitors}
%
%
%
{\speaker{V.V.~Sokolov}\inst{1}}
{%
\inst{1}%
SAO RAS%
\instaddr{, Nizhny Arkhyz, 357147, Karachai-Cherkessia, Russia \\}}
{sokolov@sao.ru}
%
%
%
{}
{}
{}{}{}


\abstext
{
The recent observational data show that collapse of massive stars
is a more preferable scenario to produce so-called long-duration GRBs, than
naive expectations on merging of binary compact objects, which (the
NS+NS scenario) appeared before the first determination of redshifts,
measurements of spectra and images. Observed location distribution of
variable optical sources, or optical transients (OT GRB), relative to their
host galaxies allows us to suppose that GRB sources are associated with
vigorous massive star-forming in distant galaxies: spiral, irregular,
blue compact and others with transient burst of star formation in them.
At least some OT GRBs can be located directly in star-forming regions (or
in their vicinities) of these galaxies in which (massive) star formation
rate is tens times higher than in the same galaxies in the local Universe.
In addition to the characteristic ``knee'' on the light curve of OT
GRB~970508 revealed by $I$-band photometric observations with the 6-m
telescope, more evidences (GRB 970228, GRB 980326, 990712, 991208) of link
between GRBs and Type Ib/c SNe (or core-collapse SNe) were found, 
which can be
an additional argument in favor of the idea of massive stars as progenitors
of cosmic gamma-ray bursts. 
The observations of K$_{\alpha}$ lines of iron
in afterglow X-ray spectra of GRBs (970508, 970828, 991216, 000214) and the
observation of redshifted absorption feature of neutral iron (7.1 keV)
simultaneously with the GRB 990705 give additional evidence in favor
of massive stars as progenitors of GRBs.
}
%
%

\item[\bf 17.50--18.10]
\contribution
%
%
{The modeling of spectra of GRB host galaxies}
%
%
%
{\speaker{T.A.~Fatkhullin}\inst{1}, V.V.~Sokolov\inst{1}}
{%
\inst{1}%
SAO RAS%
\instaddr{, Nizhny Arkhyz, 357147, Karachai-Cherkessia, Russia \\}}
{timur@sao.ru}
%
%
%
{}
{}
{}{}{}


\abstext
{
The modeling of some galaxies, where GRBs were detected, is performed.
The main parameters of these galaxies (with characteristic color and
spectral features of star formation burst): luminosity, mass, mean
star formation rate and probable star formation scenario are determined
taking into account extinction by dust and gas. It is shown that
these galaxies are usual, with higher star formation rate; they are
mainly observed in optics at redshifts about 1 and higher.
}

%
%

\item[\bf 18.10--18.30]
\contribution
%
%
{Evidence for a strong evolution of GRBs and \nl
constraints on the GRB intrinsic luminosity function \nl
with new GRB statistical data}
%
%
%
{\speaker{Ya.Yu.~Tikhomirova}\inst{2,3}, B.E.~Stern\inst{1,2,3}, R.~Svensson\inst{3}}
{%
\inst{1}%
Institute for Nuclear Research RAS,%
\instaddr{ Institute for Nuclear Research, 117312, Moscow, Russia \\}
\inst{2}%
Astro Space Centre FIAN,%
\instaddr{ 117810, Moscow, Russia \\}
\inst{3}%
Stockholm Observatory%
\instaddr{, SCFAB, Stockholm Observatory, Department of Astronomy, 
SE-106 91 Stockholm, Sweden \\}}
{jana@anubis.asc.rssi.ru}
%
%
%
{}
{}
{}{}{}


\abstext
{
We present new constraints on the evolution of GRBs sources 
and on their 
intrinsic luminosity function obtained with the recent data: the uniform
sample of 3300 long GRBs found in the BATSE continuous records and  the 
sample of 17 GRBs with measured redshifts. The latter sample has three GRBs
with a very high intrinsic peak photon flux. If we fix the bright end 
of intrinsic  luminosity distribution to these events and allow the rest of 
the distribution to vary as a broken power law we find: (i) nonevolving
population models predict too large number of apparently strong GRBs, they
are rejected at the 10$^{-5}$ level; (ii) the decline of the GRBs population
from $z=1.5$ towards the present epoch is approximately as sharp as the decline 
of the star production; (iii) the intrinsic luminosity function
behaves as $dN/dP \sim P^{-1.5}$ through at least two orders of magnitude. 
The evolution of gamma bursters at large redshifts cannot be constrained. 
The data fit is not sensitive to the cosmological parameters.
}

%
%

\item[\bf 18.30--18.40]
D.A.~Varshalovich {\it(Ioffe Institute)}. Closing

\end{list}

\normalsize
\newpage
~
\thispagestyle{empty}
\newpage

\markboth{\sl Abstracts}
         {\sl Physics of Neutron Stars -- 2001}

\section[Abstracts]{}

\vspace*{8cm}
\noindent
\centerline{\Huge\bf ABSTRACTS}
\thispagestyle{empty}
\newpage
~
\thispagestyle{empty}
\newpage



\renewcommand{\speaker}[1]{#1}
  
\renewcommand{\nl}{ \\[.3ex] }

\renewcommand{\nli}{ \\ }
            
\renewcommand{\inst}[1]{$^{#1}$}
  
\renewcommand{\instaddr}[1]{#1}
    
\renewcommand{\contribution}[9]{
  {\renewcommand{\inst}[1]{}
   \renewcommand{\nl}{}
   \subsection[#2. {\it #1}]{}
  }
  \vspace*{-6 mm}
  \noindent
  {\large\sc #1 }\\[2ex]
  {\bf #2 } \\[2ex]
  {\it #3} \\[-\baselineskip]
  {\sl E-mail:\ \ #4} \\[0ex]
  \ifthenelse{\not\equal{#5}{}}
    {\sl Web:\ \ \ \ \ \ \href{#5}{#6}}
    {\vspace{-\baselineskip}} \\
  \ifthenelse{\not\equal{#7}{}}
    {\ifthenelse{\not\equal{#8}{}}
      {\ifthenelse{\not\equal{#9}{}}
        {\sl E-print:\ \href{http://xxx.lanl.gov/abs/#7}{#7},
                       \href{http://xxx.lanl.gov/abs/#8}{#8},
                       \href{http://xxx.lanl.gov/abs/#9}{#9}}
        {\sl E-print:\ \href{http://xxx.lanl.gov/abs/#7}{#7},
                       \href{http://xxx.lanl.gov/abs/#8}{#8}}
      }
      {\sl E-print:\ \href{http://xxx.lanl.gov/abs/#7}{#7}}
    }
    {\vspace{-\baselineskip}}
}

\renewcommand{\abstext}[1]{
  \vspace{4ex}
  \rm
  #1
  \vskip 1cm
}

\renewcommand{\arxiv}[1]{
  \href{http://xxx.lanl.gov/abs/#1}{#1}
}

  
\renewcommand{\references}[1]{
\setcounter{refer}{1}   
\vspace{-2ex}
\noindent
{\bf References:}\\
\vspace{-3ex}
\begin{list}{\arabic{refer}.}%
            {\usecounter{refer}
             \leftmargin 5 mm
             \itemsep -1.5 mm}
#1  
\end{list}  
}

\contribution
%
%
{Theoretical interpretation of the X-ray pulsar 
spectra \nl consisting of several cyclotron harmonics}
%
%
%
{\speaker{A.N.~Baushev}\inst{1}, G.S.~Bisnovatyi-Kogan\inst{1}}
{%
\inst{1}%
Space Research Institute%
\instaddr{, 84/32, Profsoyuznaja st., 117810, Moscow, Russia \\}}
{abaushev@mx.iki.rssi.ru, gkogan@mx.iki.rssi.ru}
%
%
%
{}
{}
{}{}{}


\abstext
{
The spectrum of cyclotron radiation is calculated 
for anisotropically distributed relativistic electrons 
moving with  
nonrelativistic velocities across the magnetic field. 
It is shown that if such electrons
are responsible for the formation of the ``cyclotron'' line in 
the spectrum of Her X-1 then the value of its magnetic field, 
(3--6)$\cdot$ 10$^{10}$ G, which results from
this interpretation, is 
in a good agreement with some other observations and theoretical 
estimates. The case of multi-harmonical cyclotron line 
is also considered. Observations of temporal variations 
of the energy of this ``cyclotron'' line in the spectra of several 
X-ray pulsars is explained by variability of the average 
longitudinal energy of electrons, which decreases with increasing 
luminosity due to radiative deceleration of the accretion flow.
}

%
%

\vspace{10ex}

\contribution
%
%
{On the 2D structure of a thin accretion disk}
%
%
%
{\speaker{V.S.~Beskin}\inst{1}, R.Yu.~Kompaneetz\inst{2}}
{%
\inst{1}%
Lebedev Physical Institute,%
\instaddr{ 142292, Moscow, Russia \\}
\inst{2}%
Moscow Institute of Physics and Technology%
\instaddr{, Dolgoprudnyi 141700, Russia \\}}
{beskin@lpi.ru}
%
%
%
{}
{}
{}{}{}


\abstext
{
The internal structure of a thin accretion disk 
in the vicinity of a black hole inside the last stable orbit 
($R<$ 3$R_g$) is considered within the hydrodynamical version of 
the Grad-Shafranov equation. It is shown that at any way up to 
the sonic surface the flow is not radial, the flow parameters 
in the equatorial region (e.g., the poloidal velocity) differing 
from those near the disk boundary. 
}

\newpage

\contribution
%
%
{Phase transitions in stars: stability, pulsations, \nl
convective Urca-shells and pre-supernovae}
%
%
%
{\speaker{G.S.~Bisnovatyi-Kogan}\inst{1}}
{%
\inst{1}%
Space Research Institute%
\instaddr{, 84/32, Profsoyuznaja st., 117810, Moscow, Russia \\}}
{gkogan@mx.iki.rssi.ru}
%
%
%
{}
{}
{astro-ph/0004281}{}{}


\abstext
{
The problem of damping of stellar oscillations in the presence of Urca
shell is solved analytically in a plane symmetrical approximation.
Low amplitude oscillations are considered. Oscillatory pressure
perturbations induce beta reactions of the electron capture and
decay in a thin layer around the Urca shell, leading to damping
of oscillations. Due to nonlinear dependence of beta reaction
rates on the pulsation amplitude in a degenerate matter, even a
low amplitude oscillation damping follows a power-low. It is shown
that in the presence of the Urca shell the energy losses due to
neutrino emission, and the entropy increase due to non-equilibrium
beta reactions are much lower than the rate of decrease of the
energy of pulsations by excitation of short-wavelength
acoustic waves. Dissipation of the vibrational energy by the
latter process is the main source of heating of matter. Convective
motion in the presence of the Urca shell is considered, and
equations generalizing the mean free path model of convection
are derived. Convective motion is a source of both energy losses 
due to Urca reactions
in the shell, and nonequilibrium beta heating of degenerate matter.
This problem is closely related to thermal stability 
and boundary of SN type I explosions.
Only self-consistent evolutionary calculations may 
clarify the effect
of convective Urca-shell on the thermal stability 
of the pre-SN model.
}

%
%

\newpage

\contribution
%
%
{Gamma rays from pulsar winds}
%
%
%
{\speaker{S.V.~Bogovalov}\inst{1}, F.A.~Aharonian\inst{2}}
{%
\inst{1}%
Astrophysics Institute at the Moscow state Engineering Physics Institute,%
\instaddr{ Kashirskoje Shosse 31, 115409, Moscow, Russia \\}
\inst{2}%
Max-Planck-Institut f\"ur Kernphysik%
\instaddr{, Postfach 103980, D-69029, Heidelberg, Germany  \\}}
{bogoval@axpk40.mephi.ru, Felix.Aharonian@mpi-hd.mpg.de}
%
%
%
{}
{}
{astro-ph/0003157}{}{}


\abstext
{
The relativistic winds of  pulsars with average Lorentz-factor
$\gamma_W \sim$ 10$^6$ could be directly observed through its inverse Compton
(IC) $\gamma$-ray emission caused by illumination of the wind by
low-frequency photons emitted by the surface of the pulsar and its
magnetosphere.  Calculations performed for the Crab and Vela pulsars
show that even the interaction of the wind electrons with thermal emission of
the neutron star gives detectable fluxes of the gamma-ray emission.
Comparison of the observations of the Crab pulsar in gamma-rays above
0.5 TeV with calculations gives rather
important constraints on the mechanism of the wind acceleration.
To avoid overproduction of the pulsed TeV emission from Crab, the wind
should be accelerated at a distance exceeding 30 light cylinders in
conventional models of radio pulsars. The acceleration of the wind
close to the light cylinder
is possible only in the presence of
rather strong constraints on the mechanisms of 
generation of soft nonthermal radiation by the pulsar.
}

%
%

\vspace{10ex}

\contribution
%
%
{ Hard-energy emission from supernova remnants}
%
%
%
{\speaker{A.M.~Bykov}\inst{1}}
{%
\inst{1}%
Ioffe Physical Technical Institute%
\instaddr{, Politekhnicheskaya 26, St. Petersburg 194021 \\}}
{byk@astro.ioffe.rssi.ru }
%
%
%
{http://www.ioffe.rssi.ru/astro/DTA/DTA-Pub1.html}
{http:/$\!$/www.ioffe.rssi.ru/astro/DTA/DTA-Pub1.html}
{astro-ph/0010157}{}{}


\abstext
{Hard X-ray and gamma-ray emission has been detected from a number of 
galactic supernova remnants. We discuss relevant processes and models of 
hard emission production in supernova remnants simultaneously with the results 
of recent observations of SNRs obtained with CGRO, ASCA, BeppoSAX and 
XMM-Newton. Implications of the observed spectra of SNRs to test models of 
pulsar wind nebulae emission as well as the cosmic ray origin problem will 
be discussed.       
}

%
%
%
%

\newpage

\contribution
{Wolf-Rayet stars and relativistic objects}
{\speaker{A.M.~Cherepashchuk}\inst{1}}
{%
\inst{1}%
Sternberg Astronomical Institute%
\instaddr{, Universitetskij pr. 13, 119899, Moscow, Russia \\}} 
{cher@sai.msu.ru} 
{} 
{}
{astro-ph/0012512}{}{}

\abstext{
The observed properties of Wolf-Rayet stars and
relativistic objects in close binary systems are analyzed. The
final masses $M^{f}_{\rm CO}$ for the carbon-oxygen cores of WR
stars in WR+O binaries are calculated taking into account the
radial loss of matter via stellar wind, which depends on the mass
of the star. The analysis includes new data on the clumpy
structure of WR winds, which appreciably decreases the required
mass-loss rates $M_{\rm WR}$ for the WR stars. The masses
$M^{f}_{\rm CO}$ lie in the range (1--2)$M_\odot$ -- (20--44)$M_\odot$
and have a continuous distribution. The masses $M_x$ of the
relativistic objects are (1--20)$M_\odot$ and have a bimodal
distribution: the mean masses for neutron stars and black holes
are (1.35 $\pm $0.15$)M_\odot$ and (8--10)$M_\odot$, respectively, with
a gap from (2--4)$M_\odot$ in which no neutron stars or black holes
are observed in close binaries. The mean final CO-core mass is
$\overline M^f_{\rm CO}$ = (7.4--10.3)$M_{\odot}$, close to the mean
mass for the black holes. This suggests that it is not only the
mass of the progenitor that determines the nature of the
relativistic object, but other parameters as well --- rotation,
magnetic field, etc. One SB1R Wolf-Rayet binary and 11 suspected
WR+C binaries that may have low-mass X-ray binaries with neutron
stars and black holes.
}

\vspace{10ex}

\contribution
%
%
{Photon-neu\-trino processes in a strongly \nl
magnetized plasma}
%
%
%
{\speaker{M.V.~Chistyakov}\inst{1}, N.V.~Mikheev\inst{1}}
{%
\inst{1}%
Yaroslavl State University%
\instaddr{, Sovietskaya 13, 150000, Yaroslavl, Russia \\}}
{mch@univ.uniyar.ac.ru}
%
%
%
{}
{}
{hep-ph/9907345}{}{}


\abstext
{
The effect of strongly magnetized electron-positron plasma on
photon-neutrino processes $\nu \to \nu \gamma$ and $\gamma \nu
\to \nu $ is investigated. The amplitudes and 
probabilities of the processes are calculated taking into account 
the photon dispersion and large radiative corrections near the
resonance. It is shown that the combined effect of plasma and
strong magnetic field decreases the probability of the process
$\nu \to \nu \gamma$  in comparison with the value obtained 
neglecting the plasma effect.
The probability of the process $\gamma \nu \to \nu$ does not
depend on the initial neutrino energy and is negligibly small in
comparison with probability $W_{\nu \to \nu \gamma}$ in the low
temperature limit.
}

%
%

\newpage

\contribution
%
%
{ Separation of millisecond and submillisecond 
pulsations \nl in hard gamma radiation 
from the Geminga pulsar}
%
%
%
{\speaker{E.I.~Chuikin}\inst{1}}
{%
\inst{1}%
Ioffe Physical Technical Institute%
\instaddr{, Politekhnicheskaya 26, St. Petersburg 194021 \\}}
{evgeniy.chuikin@pop.ioffe.rssi.ru }
%
%
%
{}
{}
{}{}{}


\abstext
{
Separation of millisecond and submillisecond pulsations in radiation from 
the Geminga pulsar
(PSR J0633+1746) 
with ``Discrete temporal Fourier 
series analysis method'' on the data base of GAMMA-1 observations (from 30 
November 1990 to 2 February 1991) is discussed. The pulsations are analysed 
also with a "Superposition of epoch method".
Significant pulsations were found within the first peak of the Geminga
light curve. Among the strongest ones there 
are the pulsations with period values
6.1313 ms, 1.7060 ms, 0.3966 ms. 
As in the previous studies of pulsations of the Vela pulsar,
reliability of the results obtained for Geminga was confirmed by 
the phasogram pattern and phase of the peak being stable
throughout the observational set.
}

%
%
%
%

\vspace{10ex}

\contribution
%
%
{On the competition between synchrotron and Compton \nl
emission of electrons in relativistic shocks}
%
%
%
{\speaker{E.V.~Derishev}\inst{1}, V.V.~Kocharovsky\inst{1}, Vl.V.~Kocharovsky\inst{1}}
{%
\inst{1}%
Institute of Applied Physics%
\instaddr{, 46 Ulyanov st., 603600, Nizhny Novgorod, Russia \\}}
{derishev@appl.sci-nnov.ru}
%
%
%
{}
{}
{}{}{}


\abstext
{
We develop a self-consistent theory of the 
synchrotron-self-Compton 
emission for optically thin sources with a stationary injection of 
relativistic mono-energetic electrons. We investigate the electron 
distribution function, synchrotron and inverse Compton (IC) spectra 
and find their analytical asymptotes. It is shown that
self-consistent steady-state 
electron distribution may produce a variety of low-energy spectral 
indices ranging from 1/2 to 1. 

The spectrum of comptonized radiation is entangled with low-energy 
synchrotron emission via the electron distribution. We derive explicit 
expression allowing one to find IC spectrum by means of the integral 
transformation of synchrotron spectrum.

Applying this theory to Gamma-Ray Burst sources we find new 
arguments in favor of synchrotron origin of the observed sub-MeV 
radiation.
}

%
%

\newpage

\contribution
%
%
{Self-sustained neutron haloes in the inner parts \nl of hot accretion disks}
%
%
%
{\speaker{E.V.~Derishev}\inst{1}, A.A.~Belyanin\inst{1}}
{%
\inst{1}%
Institute of Applied Physics%
\instaddr{, 46 Ulyanov st., 603600, Nizhny Novgorod, Russia \\}}
{derishev@appl.sci-nnov.ru}
%
%
%
{}
{}
{}{}{}


\abstext
{
We suggest and analyze in detail the possibility of self-sustained  
neutron halo formation in the vicinity of disk-accreting black hole 
or neutron star. Initial seed neutrons originate from collisional 
dissociation of helium in a hot infalling plasma. Once the neutron halo 
is formed, it becomes ``self-sustained'', i.e.,  supported by 
collisions of energetic neutrons from the halo with helium nuclei 
even if  the  ion temperature in the disk drops below the threshold  
for He dissociation by protons. 
 
Our analysis shows that the halo formation is possible when the mass  
accretion rate is below the critical value $\dot{M}_{max} \sim$  
5$\cdot$ 10$^{16}M/M_{\odot}$~g/s, where $M$ is the black-hole mass. 
In this case neutrons accrete slower than ions which leads to the  
neutron pile-up: the density of neutrons can be much higher than the  
proton density. Under certain conditions neutrons may diffuse 
outwards into the cooler part of an accretion disk, leading to 
enhanced deuterium production accompanied by emission in a relatively 
narrow 2.2 MeV line.
 
Neutron halo exists in a broad range of disk parameters and mass 
accretion rates. We discuss the observability of various dynamical 
and radiative phenomena related to the existence of neutron halo.
}

%
%

\newpage

\contribution
%
%
{The modeling of spectra of GRB host galaxies}
%
%
%
{\speaker{T.A.~Fatkhullin}\inst{1}, V.V.~Sokolov\inst{1}}
{%
\inst{1}%
SAO RAS%
\instaddr{, Nizhny Arkhyz, 357147, Karachai-Cherkessia, Russia \\}}
{timur@sao.ru}
%
%
%
{}
{}
{}{}{}


\abstext
{
The modeling of some galaxies, where GRBs were detected, is performed.
The main parameters of these galaxies (with characteristic color and
spectral features of star formation burst): luminosity, mass, mean
star formation rate and probable star formation scenario are determined
taking into account extinction by dust and gas. It is shown that
these galaxies are usual, with higher star formation rate; they are
mainly observed in optics at redshifts about 1 and higher.
}

%
%

\vspace{10ex}

\contribution
%
%
{Short gamma ray bursts}
%
%
%
{\speaker{D.D.~Frederiks}\inst{1}}
{%
\inst{1}%
Ioffe Physical Technical Institute%
\instaddr{, Politekhnicheskaya 26, St. Petersburg 194021 \\}}
{fred@pop.ioffe.rssi.ru }
%
%
%
{}
{}
{}{}{}


\abstext
{
A review of ``short GRB'' observations with Konus instruments is given.
An observational base of this intriguing class of bursts contains now
more than 100 events. General properties of short GRBs are considered in
comparison with long events. Some interesting bursts are discussed
in detail.
}

%
%

\newpage

\contribution
%
%
{The origin of QPOs of X-ray binaries}
%
%
%
{\speaker{Yu.N.~Gnedin}\inst{1}, S.O.~Kiikov\inst{1}}
{%
\inst{1}%
Central (Pulkovo) Astronomical Observatory%
\instaddr{, Pulkovskoe Shosse 65/1, 196140, \nl St.-Petersburg, Russia \\}}
{gnedin@gao.spb.ru, kiikov@gao.spb.ru}
%
%
%
{}
{}
{}{}{}


\abstext
{We present three mechanisms of generation of X-ray quasi-periodic 
oscillations (QPOs) in binaries. Two of them are based on the analogy 
to nonlinear oscillations of gaseous caverns in liquid. The first 
mechanism is called magnetocavitation;
it implies that X-ray QPOs of a neutron 
star are produced by radial oscillations of the neutron star magnetosphere 
interacting with accreting plasma. To study X-ray QPOs of neutron stars 
with critical (Eddington) luminosity, the photon-cavitation mechanism has 
been considered. In this case X-ray QPOs of the neutron star are generated 
by radial oscillations of photon caverns in a fully ionized hydrogen 
plasma settling in the accreting column of the compact object. To explain 
X-ray QPOs of binaries with a black hole and cataclysmic variables we suggest 
the mechanism according to which X-ray QPOs are caused by nonlinear 
oscillations of current sheets originating in accreting disks. The calculated 
values of basic physical parameters of QPOs such as basic frequencies, 
frequency and amplitude dependence on the X-ray flux level and the energy of 
photons and also on QPOs lag time of photons of different energies are 
consistent with the observational data.
}

%
%

\vspace{10ex}

\contribution
%
%
{Magnetocavitation mechanism of gamma-ray burst \nl phenomenon}
%
%
%
{\speaker{Yu.N.~Gnedin}\inst{1}, S.O.~Kiikov\inst{1}}
{%
\inst{1}%
Central (Pulkovo) Astronomical Observatory%
\instaddr{, Pulkovskoe Shosse 65/1, 196140, \nl St.-Petersburg, Russia \\}}
{gnedin@gao.spb.ru, kiikov@gao.spb.ru}
%
%
%
{}
{}
{astro-ph/9908124}{}{}


\abstext
{We consider the phenomenon of a gamma-ray burst as a nonlinear collapse 
of a magnetic cavity surrounding a neutron star with huge magnetic 
field due to the bubble shape instability in a resonant 
MHD field of an accreting plasma or a plasma
on the neutron star surface. The QED 
effect of vacuum polarizability by a strong magnetic field is taken into 
consideration. We develop the analogy with the phenomenon of 
sonoluminescence in which the gas bubble is located in surrounding liquid 
with a driven sound intensity. 
}

%
%

\newpage

\contribution
%
%
{X-ray bursts with signs of strong photospheric radius \nl
expansion of a neutron star}
%
%
%
{\speaker{S.A.~Grebenev}\inst{1}, S.V.~Molkov\inst{1}, 
A.A. Lutovinov\inst{1}, R.A. Sunyaev\inst{1}}
{%
\inst{1}%
Space Research Institute%
\instaddr{, 84/32, Profsoyuznaja st., 117810, Moscow, Russia \\}}
{sergei@hea.iki.rssi.ru}
%
%
%
{}
{}
{astro-ph/0005082}{}{}


\abstext
{
We present results of the GRANAT/ART-P and RXTE/PCA observations of
several X-ray bursts with obvious signs of strong photospheric radius
expansion (precursors and dips in profiles, delayed rise phase, etc).
In particular, the photospheric radius exceeded 70 km during an
intense X-ray burst detected from the source 1E1724-307 in the
globular cluster Terzan 2. After the precursor which was comparable
in strength with the main event the flux from the source fell below
the persistent level. The very narrow precursor was detected in the
profile of a long ($\sim$ 10 min) X-ray burst observed from GX17+2. We
analyze correlations in the effective temperature and radius of the
photosphere in these and other X-ray bursters and note that
Comptonization may be responsible for the detected features. To
describe in detail spectral evolution of the source during the burst
a number of model spectra were computed taking into account both
Comptonization and free-free absorption in the outer layers of the
photosphere.
}

%
%

\vspace{10ex}

\contribution
%
%
{Efficiency of $e^+e^-$ pair production by neutrinos} 
%
%
%
{\speaker{A.A.~Gvozdev}\inst{1}, I.S.~Ognev\inst{1}}
{%
\inst{1}%
Yaroslavl State University%
\instaddr{, Sovietskaya 13, 150000, Yaroslavl, Russia \\}}
{gvozdev@univ.uniyar.ac.ru} 
%
%
%
{}
{}
{}{}{}


\abstext
{
The dominating processes of the neutrino production and
$e^+ e^-$-pair creation by such neutrinos are 
investigated in the model of hyper-accreting disk around 
a Kerr black hole. These processes are studied both 
in vacuum  and  in the presence of strong magnetic 
field. It is shown that the efficiency of $e^+ e^-$-plasma 
production is too small to explain the 
origin of cosmological GRBs.
}

%
%

\newpage

\contribution
%
%
{Unified equations of state of dense matter and \nl neutron star structure }
%
%
%
{\speaker{P. Haensel}\inst{1}}
{%
\inst{1}%
N. Copernicus Astronomical Center%
\instaddr{, Bartycka 18, 00-716 Warszawa, Poland\\}}
{haensel@camk.edu.pl}
%
%
%
{}
{}
{astro-ph/0006135}{astro-ph/0105485}{}


\abstext
{
Unified equation of state (EOS)  describes in a physically 
consistent way neutron star matter within the crust and the liquid 
core, and is based on a single nuclear hamiltonian, ${\hat H}_{\rm N}$. 
To make a solution of the nuclear many-body problem feasible, one uses 
an effective nuclear hamiltonian, ${\hat H}^{\rm eff}_{\rm N}$. It has 
to reproduce the ground state properties of laboratory nuclei, in
particular of those with a large neutron excess, and to 
give correct saturation 
parameters of nuclear matter. Additional conditions, relevant for correct
description of very neutron-rich matter in neutron stars have to be also 
imposed on ${\hat H}_{\rm N}^{\rm eff}$. 
Two examples of  ${\hat H}_{\rm N}^{\rm eff}$ are: an older 
FPS ({\bf F}riedman-{\bf P}andharipande-{\bf S}kyrme) 
 model (Pandharipande \& Ravenhall 1989), and a recent SLy 
({\bf S}kyrme-{\bf L}yon) one 
(Chabanat et al. 1997, 1998). Corresponding unified EOSs were  
calculated, assuming ground state, $T$=0, for  the crust matter, 
and  simplest (minimal) $npe\mu$ composition of the 
liquid core: FPS EOS (Lorenz et al. 1993), SLy EOS (Douchin \& Haensel 2000, 
2001). 
 Properties of the bottom layers of the neutron star 
crust will be  discussed, with particular emphasis on the crust core 
interface (Douchin et al., Douchin \& Haensel 2000). 
 Neutron star models, calculated using  these EOSs (Douchin \& Haensel 2001), 
 will be  reviewed. In particular, properties of 
equilibrium configurations  near minimum mass  (Haensel, Zdunik, 
Douchin 2001), will be  studied, and related 
to the specific features of the EOS near crust-core interface.  
 Effects 
of rotation on both $M_{\rm max}$ and $M_{\rm min}$, and on the 
mass-radius curves  
(Haensel, Zdunik, Douchin 2001) 
 will be 
briefly  reviewed.  
Finally, parameters of neutron star models will be  confronted with available 
observational data (Douchin \& Haensel 2001, Haensel 2001). 
}

%
%
\references{
\item
E.~Chabanat, P.~Bonche, P.~Haensel,  J.~Meyer, R.~Schaeffer. 
Nucl. Phys. A 627, 710 (1997)
\item
E.~Chabanat, P.~Bonche, P.~Haensel,  J.~Meyer, R.~Schaeffer. 
Nucl. Phys. A 635, 231  (1998)
\item
F.~Douchin, P.~Haensel, J.~Meyer. Nucl. Phys. A 665, 419 (2000)
\item
F.~Douchin, P.~Haensel. Phys. Lett. B 485, 107 (2000)
\item
F.~Douchin, P.~Haensel, J.~Meyer. Nucl. Phys. A 665, 419 (2000)
\item
P.~Haensel. Apparent radii of neutron stars, observations of Geminga and 
RX J185635-3754, and equation of state dense matter. A\&A Letters
(subm., \arxiv{astro-ph/0105485})
\item
P.~Haensel, J.L.~Zdunik, F. Douchin. Equation of state of dense matter and the 
minimum mass of cold neutron stars, in preparation (2001)
\item
C.P. Lorenz, D.G. Ravenhall, C.J.~Pethick. Phys. Rev. Lett. 70, 379 (1993)
\item
V.R. Pandharipande, D.G. Ravenhall. In: M. Soyeur et al. (Eds.), Proc. NATO 
Advanced  Research Workshop  on nuclear matter and heavy ion 
collisions, Les Houches, Pleanum, New York (1989)
}

\newpage

\contribution
%
%
{Gravitational wave background from coalescing compact \nl
stars in eccentric orbits}
%
%
%
{\speaker{V.B.~Ignatiev}\inst{1}, A.G.~Kuranov\inst{2}, K.A.~Postnov\inst{2},
M.E.~Prokhorov\inst{2}}
{%
\inst{1}%
Faculty of Physics, Moscow State University,%
\instaddr{ 119899 Moscow, Russia \\}
\inst{2}%
Sternberg Astronomical Institute%
\instaddr{, Universitetskij pr. 13, 119899, Moscow, Russia \\}}
{whirl@xray.sai.msu.ru}
%
%
%
{}
{}
{astro-ph/0106299}{}{}


\abstext
{
Stochastic gravitational wave background produced by a 
stationary coalescing population of binary neutron 
stars in the Galaxy is calculated. This background is 
found to constitute a confusion limit within the LISA 
frequency band up to a limiting frequency 
$\nu_{lim}\sim $10$^{-3}$~Hz, leaving the frequency 
window $\sim $10$^{-3}$--10$^{-2}$~Hz open for
potential detection of cosmological stochastic 
gravitational waves.
}

\vspace{2ex}

\contribution
%
%
{Masses of
isolated neutron stars and superfluid gaps \nl
in their cores: constraints from observations and \nl
cooling theory
}
%
%
%
{\speaker{A.D.~Kaminker}\inst{1}, P.~Haensel\inst{2}, D.G.~Yakovlev\inst{1}}
{%
\inst{1}%
Ioffe Physical Technical Institute,%
\instaddr{ Politekhnicheskaya 26, St. Petersburg 194021 \\}
\inst{2}%
N.~Copernicus Astronomical Center%
\instaddr{, Bartycka 18, 00-716 Warszawa, Poland\\}}
{kam@astro.ioffe.rssi.ru}
%
%
%
{http://www.ioffe.rssi.ru/astro/NSG/NSG-Pub1.html}
{http:/$\!$/www.ioffe.rssi.ru/astro/NSG/NSG-Pub1.html}
{astro-ph/0105047}{}{}


\abstext
{
We present the results of new cooling simulations
of neutron stars under the assumption
that stellar cores consist of neutrons, protons
and electrons. We assume further that
nucleons may be in a superfluid state and 
use realistic density profiles
of superfluid critical temperatures
$T_{\rm cn}(\rho)$ and $T_{\rm cp}(\rho)$ of neutrons and protons.
Taking a suitable profile of $T_{\rm cp}(\rho)$ 
with maximum $\sim 5 \times 10^9$ K we
obtain smooth transition from slow to rapid cooling
with increasing stellar mass. Adopting
the same profile we can explain the majority of observations of thermal 
emission from isolated middle--aged
neutron stars by cooling
of neutron stars with different masses  either
with no neutron superfluidity in the cores
or with a weak superfluidity, $T_{\rm cn} < 10^8$ K.
The required masses depend sensitively
on the decreasing slope of the $T_{cp}(\rho)$ profile
at $\rho \sim 10^{15}$ g cm$^{-3}$. 
For one particular model $T_{cp}(\rho)$ profile the masses range 
from $\sim 
1.2 \, {\rm M}_\odot$ for
(young and hot) RX J0822--43 and 
(old and warm) PSR 1055--52 
and RX J1856-3754
to $\approx 1.45\, {\rm M}_\odot$
for the (rather cold) Geminga and Vela pulsars.
Shifting the decreasing slope of $T_{cp}(\rho)$
profile to higher densities
we may obtain higher masses of the same sources.
This gives a new method to constrain neutron star
masses and superfluid critical temperatures in the stellar
cores.
}

%
%
\references{

\item
Kaminker A.D., Haensel P., Yakovlev D.G. 2001, A\&A Lett., in press
(\arxiv{astro-ph/0105047})

}

\newpage

\contribution
%
%
{Interaction of anisotropic pulsar winds \nl 
with interstellar medium
}
%
%
%
{\speaker{D.V.~Khangoulian}\inst{1}, S.V.~Bogovalov\inst{1}}
{%
\inst{1}%
Astrophysics Institute at the Moscow state Engineering Physics Institute%
\instaddr{, Kashirskoje Shosse 31, 115409, Moscow, Russia \\}}
{bogoval@axpk40.mephi.ru}
%
%
%
{}
{}
{}{}{}


\abstext
{
The objective of the work is to explain the 
morphology of X-ray plerions around the Crab and the Vela pulsars observed 
with the Chandra telescope. The X-ray plerions consist of toroidal structure and 
jet-like features directed along the axis of rotation. We assume that the 
observed structures are produced 
due to interaction of the anisotropic relativistic winds ejected by radio 
pulsars with interstellar medium. MHD models of the wind formation predict 
that the energy flux in the wind is concentrated at the equatorial region.  
For one of the simplest distributions of the energy flux in the wind
we have calculated the shape of the terminating shock wave and 
compare it with observations.
}

%
%

\vspace{10ex}

\contribution
%
%
{Multiband photometry of Geminga}
%
%
%
{\speaker{V.N.~Komarova}\inst{1}, V.G.~Kurt\inst{2}, T.A.~Fatkhullin\inst{1},
V.V.~Sokolov\inst{1}, Yu.A.~Shibanov\inst{3}, A.B.~Koptsevich\inst{3}}
{%
\inst{1}%
SAO RAS,%
\instaddr{ Nizhny Arkhyz, 357147, Karachai-Cherkessia, Russia  \\}
\inst{2}%
Astro Space Centre FIAN,%
\instaddr{ 117810, Moscow, Russia \\}
\inst{3}%
Ioffe Physical Technical Institute%
\instaddr{, Politekhnicheskaya 26, St. Petersburg 194021 \\}}
{vkom@sao.ru}
%
%
%
{}
{}
{}{}{}


\abstext
{
The results of Geminga's investigation with the aid of 
BVRI-photometry based on observations with the 6\,m 
telescope are presented. The obtained Geminga's      
magnitudes B (26.1$\pm$0.5), V (25.3$\pm$0.3) and 
R$_c$ (25.4$\pm$0.3) are consistent with the results of 
other studies of its optical emission in this spectral 
range. We derive for the first time the magnitude in 
I$_c$-band to be $25^m.1\pm0.4$, which is more than 
a magnitude higher if compared with the upper limit 
given in the paper by Bignami et al. (1996).                   
The comparison of the broadband spectra of this 
middle-aged isolated neutron star with the results of 
observations in X-rays and hard ultraviolet allows us
to assume the presence of the features probably of nonthermal 
character in Geminga's emission at least in some 
regions of the visible spectrum.                    
}

%
%

\newpage

\contribution
%
%
{GRB as the result of interaction of supernova ejecta \nl
with NS companion of a closed binary
}
%
%
%
{\speaker{B.V.~Komberg}\inst{1}, Ya.N.~Istomin\inst{2}}
{%
\inst{1}%
Astro Space Centre FIAN,%
\instaddr{ 117810, Moscow, Russia \\}
\inst{2}%
Lebedev Physical Institute%
\instaddr{, 142292, Moscow, Russia \\}}
{bkomberg@ASC.rssi.ru}
%
%
%
{}
{}
{}{}{}


\abstext
{
The supernova explosion (I b/c or I bw  type) in a closed binary
(with separation of 10$^{13}$ cm) can produce
magnetospheric flare possessing the properties of a GRB. According
to our estimates, the
NS magnetosphere intercepts 10$^{-11}$ of the full kinetic energy
of the blast wave (10$^{47}$ ergs). Extended magnetospheric tail 
(10$^{10}$ cm)
with the mean magnetic field 10$^6$ Gauss is the source of 
gamma-rays ejected into a
small solid angle (0.1 rad). This radiation has the synchrotron nature 
being produced by accelerated particles ($E\sim1$ Gev). Fast electrons appear
as a result of the magnetic field reconnection in the current layer,
directed along the shock velocity. Such a model explains naturally 
high anisotropy of gamma-ray emission and connection of GRBs with
phenomena of compact SNe.
}

%
%

\vspace{10ex}

\contribution
%
%
{First detection of the Vela pulsar in IR with the VLT}
%
%
%
{\speaker{A.B.~Koptsevich}\inst{1}, P.~Lundqvist\inst{2},
J.~Sollerman\inst{3}, Yu.A.~Shibanov\inst{1}, S.~Wagner\inst{4}}
{%
\inst{1}%
Ioffe Physical Technical Institute,%
\instaddr{ Politekhnicheskaya 26, St. Petersburg 194021 \\}
\inst{2}%
Stockholm Observatory,%
\instaddr{ SCFAB, Stockholm Observatory, Department of Astronomy, 
SE-106 91 Stockholm, Sweden \\}
\inst{3}%
ESO,%
\instaddr{ Karl-Schwarzschild Strasse 2, D-85748, Garching bei M\"unchen,
Germany \\}
\inst{4}%
University of Heidelberg%
\instaddr{, Landessternwarte K\"onigstuhl, D-69117, Heidelberg, Germany \\}}
{kopts@astro.ioffe.rssi.ru}
%
%
%
{}
{}
{}{}{}


\abstext
{
We report detection of the Vela pulsar 
in near-infrared during observation 
performed with VLT/ISAAC on December 2000 -- January 2001.
The pulsar is clearly identified at the images in J and H bands.
Preliminary estimations show that the  emission  
is of nonthermal origin. By this detection Vela 
starts to be a forth member of a small family 
of radio pulsars (Crab, Geminga,  and PSR B0656+14)   
detected in infrared.  
}

%
%

\newpage

\contribution
%
%
{Multiband photometry of the PSR B0656+14 \nl and its neighborhood}
%
%
%
{\speaker{A.B.~Koptsevich}\inst{1}, G.G.~Pavlov\inst{2},
S.V.~Zharikov\inst{3}, V.V.~Sokolov\inst{4}, Yu.A.~Shibanov\inst{1}, V.G.~Kurt\inst{5}}
{%
\inst{1}%
Ioffe Physical Technical Institute,%
\instaddr{ Politekhnicheskaya 26, St. Petersburg 194021 \\}
\inst{2}%
The Pennsylvania State University,%
\instaddr{ Department of Astronomy \& Astrophysics, 525 Davey Lab,
University Park, PA 16802, USA; \\}
\inst{3}%
Obs. Astr. Nacional. de Inst. de Astronomia de UNAM,%
\instaddr{ Observatorio Astronomico Nacional de Instituto de Astronomia de UNAM,
Ensenada,  B.C., 228, Mexico \\}
\inst{4}%
SAO RAS,%
\instaddr{ Nizhny Arkhyz, 357147, Karachai-Cherkessia, Russia \\}
\inst{5}%
Astro Space Centre FIAN%
\instaddr{, 117810, Moscow, Russia \\}}
{kopts@astro.ioffe.rssi.ru}
%
%
%
{http://www.ioffe.rssi.ru/astro/NSG/obs/0656-phot.html}
{http:/$\!$/www.ioffe.rssi.ru/astro/NSG/obs/0656-phot.html}
{astro-ph/0009064}{}{}


\abstext
{
We present the results of broad-band photometry
of the nearby middle-aged radio pulsar PSR B0656+14 and its neighborhood
obtained with the 6-meter telescope of the SAO RAS and with the {\sl
Hubble Space Telescope}.
The broad-band
spectral flux $F_\nu$
of the pulsar decreases with increasing frequency in the near-IR
range and increases with frequency in the near-UV range.
The increase towards UV can
be naturally interpreted as
the Rayleigh-Jeans tail of the soft thermal
component of the X-ray spectrum emitted
from the
surface of the cooling neutron star.
Continuation of the power-law component, which dominates
in the
high-energy tail of the
X-ray spectrum, to the IR-optical-UV frequencies is consistent with the
observed fluxes.
This suggests that
the non-thermal pulsar radiation may be of the same origin
in a broad frequency range from IR to hard X-rays.
We also studied 4 objects detected within
5" from the pulsar.
}

\references{
\item
Koptsevich A.B., Pavlov G.G., Zharikov S.V., Sokolov V.V., Shibanov Yu.A., Kurt V.G.
2001, A\&A, 370, 1004
}

\vspace{2ex}

\contribution
%
%
{X-ray emission from young supernova remnants}
%
%
%
{\speaker{D.I.~Kosenko}\inst{1}, S.I.~Blinnikov\inst{1,2},
E.I.~Sorokina\inst{1}, K.A.~Postnov\inst{1}}
{%
\inst{1}%
Sternberg Astronomical Institute,%
\instaddr{ Universitetskij pr. 13, 119899, Moscow, Russia \\} 
\inst{2}%
Institute for Theoretical and Experimental Physics%
\instaddr{, 117259, Moscow, Russia \\}}
{lisett@xray.sai.msu.ru}
%
%
%
{}
{}
{}{}{}


\abstext
{
We present the results of 
hydrodynamical simulations of the Tycho SN remnant. Our
model is one-dimensional and spherically symmetrical but
it takes into account
kinetics and various types of radiative transfer. 
We can obtain detailed theoretical X-ray luminosity
profiles of the remnant in different lines (silicon and iron) and
compare it with the observational results. There are several models of
type Ia SNe with different abundances and density profiles, so we could make an
assumption about progenitor of the Tiho SN and choose the most viable
model.
}

%
%

\newpage

\contribution
%
%
{A model of sub-Eddington accretion onto a magnetized \nl neutron star}
%
%
%
{\speaker{A.M.~Krassilchtchikov}\inst{1}, A.M.~Bykov\inst{1}}
{%
\inst{1}%
Ioffe Physical Technical Institute%
\instaddr{, Politekhnicheskaya 26, St. Petersburg 194021 \\}}
{kra@astro.ioffe.rssi.ru}
%
%
%
{}
{}
{}{}{}


\abstext
{
A non-stationary one-dimentional collisionless
two-fluid model of sub-Eddington accretion onto a
magnetized neutron star is developed to be used
for modeling of hard emission from X-ray binaries.
        Temporal evolution of accreting flows is
studied for a range of accretion rates and magnetic
fields within a first-order Godunov scheme with source
terms splitting.
        Strong shocks accompanied by hot plasma
regions are found to develop on timescales of about
10$^{-5}$ s; they are stable up to 10$^{-2}$ s or even longer.
        Hard emission from the hot regions may be
detected by modern X-ray and gamma-ray missions.
In this way parametres of the model may be constrained
and typical physical conditions in the flow revealed.
}

%
%

\vspace{10ex}

\contribution
%
%
{The evolution of binary stars in a globular cluster}
%
%
%
{\speaker{A.G.~Kuranov}\inst{1}, K.A.~Postnov\inst{2},
M.E.~Prokhorov\inst{2}}
{%
\inst{1}%
Faculty of Physics, Moscow State University,%
\instaddr{ 119899 Moscow, Russia \\}
\inst{2}%
Sternberg Astronomical Institute%
\instaddr{, Universitetskij pr. 13, 119899, Moscow, Russia \\}}
{alex@xray.sai.msu.ru}
%
%
%
{}
{}
{}{}{}


\abstext
{
We  present a model for the 
evolution of binary populations in a globular
cluster. The evolution of a population of close binaries is traced in a
collisional environment of evolving single stars. We consider primordial
binaries, formed simultaneously with the single stars, as well as tidally
captured binaries formed from encounters between stars. Any binary evolves
due to internal processes (e.g. evolution of its components, stellar winds,
mass transfer, etc.) and due to encounters with single stars of the
clusters. We trace individual histories of all binaries under the action 
of different physical processes, such as mass segregation,
scattering recoil, escape from the cluster. The results of our calculations
can be applied to study formation and evolution of LMXBs and millisecond
pulsars in globulars.        	
}

%
%

\label{kuranov}
\newpage

\contribution
%
%
{Comparative analysis of radio luminosity \nl
of millisecond and normal pulsars }
%
%
{\speaker{A.D.~Kuzmin}\inst{1}}
{%
\inst{1}%
Pushchino Radio Astronomy Observatory%
\instaddr{, Pushchino, 142292, Moscow Region, Russia \\}}
{akuzmin@prao.psn.ru}%
%
%
{}
{}
{}{}{}


\abstext {
We present the results of 
comparative analysis of
the integral radio luminosity
of the millisecond and normal pulsars.

Analysis is based on our measurements of the flux densities,
spectra and integral luminosities of 30 millisecond pulsars and
on the data, borrowed from the literature, which allows us to construct
the integral radio luminosities of 485 ``normal'' pulsars.

We find that contrary to the great difference of millisecond and
``normal'' pulsars in spin periods $P$, period derivatives $\dot P$,
magnetic field strengths $B$, and characteristic ages $\tau$, the
integral radio luminosities of these two pulsar populations are
nearly equal. The same is true for their dependences
on $P, \dot P, B, \tau$ and on losses of the kinetic energy $\dot E$.

We find that the integral luminosities of both millisecond and
``normal'' populations of pulsars are nearly proportional to the
parameter $B/P^2$, which characterizes potential difference
between the base and the top of the gap of a polar cap.

We suggest that millisecond and ``normal'' pulsars have similar
mechanism of radio emission, in which energetic properties are
controlled by the potential difference between the base and the
top of the gap of the polar cap.}

%
\references{

\item
A.D. Kuzmin, B.Ya. Losovsky. A\&A 368, 230 (2001)

\item
J.H. Taylor, R.N. Manchester, A.G. Lyne, F. Camilo, unpublished
 (1995)
}

\newpage

\contribution
{Three-vertex loop processes in strong magnetic field}
{\speaker{A.V.~Kuznetsov}\inst{1}, M.V.~Chistyakov\inst{1}, N.V.~Mikheev\inst{1}}
{%
\inst{1}%
Division of Theoretical Physics, Yaroslavl State University%
\instaddr{, Sovietskaya 13, 150000, \nl Yaroslavl, Russia \\}}
{avkuzn@uniyar.ac.ru}
{}
{}
{hep-ph/9808246}{hep-ph/9804444}{}

\abstext
{
A general analysis of the three-vertex loop amplitude in
very strong magnetic field, based on the asymptotic form of 
the electron propagator in the strong field, is performed.
In order to investigate the photon-neutrino process
$\gamma \gamma \to \nu \bar\nu $ and the photon splitting
$\gamma \to \gamma \gamma$, the vertex combinations of the
scalar--vector--vector ($SVV$), pseudoscalar--vector--vector ($PVV$),
3--vector ($VVV$), and axial-vector--vector--vector ($AVV$)
types are considered. It is shown that only the $SVV$ amplitude grows
linearly with the magnetic field strength, while in the
other amplitudes, $PVV$, $VVV$, and $AVV$, the linearly growing
terms are exactly canceled.
The process $\gamma \gamma \to \nu \bar\nu$ is investigated 
in the left-right-symmetric extension of the standard model 
of electroweak interaction, where the effective scalar $\nu \nu e e$
coupling is possible. 
Using the $VVV$ amplitude, the process of the photon
splitting $\gamma \to \gamma \gamma$ is investigated both below and above
the pair creation threshold.
The splitting probability is calculated taking account of the photon
dispersion and large radiative corrections near the resonance.
Possible astrophysical manifestations of the considered
processes are discussed.
Some previous results were published in the references below.
}

\references{

\item
Phys. Lett. B {\bf 434}, 67 (1998) (\arxiv{hep-ph/9804444}). 

\item
Yad. Fiz. {\bf 62}, 1638 (1999)
[Phys. At. Nucl. {\bf 62}, 1535 (1999)].

\item
In: {\it Proc. Ringberg Euroconf. "New Trends in Neutrino Physics"},
Ed. by B. Kniehl et al., 
World Sci. (1999), p. 245 (\arxiv{hep-ph/9808246}).

}

\vspace{-1ex}

\contribution
%
%
{Viscous damping of neutron star pulsations}
%
%
%
{\speaker{K.P.~Levenfish}\inst{1}, P.~Haensel\inst{2}, D.G.~Yakovlev\inst{1}}
{%
\inst{1}%
Ioffe Physical Technical Institute,%
\instaddr{ Politekhnicheskaya 26, St. Petersburg 194021 \\}
\inst{2}%
N. Copernicus Astronomical Center%
\instaddr{, Bartycka 18, 00-716 Warszawa, Poland\\}}
{ksen@astro.ioffe.rssi.ru}
%
%
%
{http://www.ioffe.rssi.ru/astro/NSG/NSG-Pub1.html}
{http:/$\!$/www.ioffe.rssi.ru/astro/NSG/NSG-Pub1.html}
{astro-ph/0004183}{astro-ph/0103290}{}


\abstext
{
Bulk and shear viscosities of matter in neutron star
cores composed of nucleons, hyperons and/or quarks
are discussed taking into account the effects of
superfluidity of various particle species. 
The bulk viscosity of nonsuperfluid matter
is enhanced by 4--5 orders of magnitude
if direct Urca process of neutrino emission is open,
and it is further enhanced by several orders of
magnitude in the presence of 
hyperons and quarks. Strong superfluidity of
matter may greatly dump the bulk viscosity.
Viscous damping times of neutron star pulsations
(particularly, r-modes) are shown to be very sensitive
to composition and superfluid properties of dense matter.
In particular, this is crucial for gravitational radiation
driven instabilities.
}





\newpage

\contribution
%
%
{Review of X-ray bursters in the Galactic Center Region}
%
%
%
{\speaker{A.A.~Lutovinov}\inst{1}, S.A.~Grebenev\inst{1}, S.V.~Molkov\inst{1}}
{%
\inst{1}%
Space Research Institute%
\instaddr{, 84/32, Profsoyuznaja st., 117810, Moscow, Russia \\}}
{aal@hea.iki.rssi.ru}
%
%
%
{}
{}
{astro-ph/0009349}{astro-ph/0105001}{astro-ph/0105002}


\abstext
{
Results of observations of X-ray bursters in the Galactic Center
region carried out with the RXTE observatory and the ART-P telescope
on board GRANAT are presented. Eight X-ray bursters (A1742-294,
SLX1744-299/300, GX3+1, GX354-0, SLX1732-304, 4U1724-307, KS1731-260)
were studied in this region during five series of observations which
were performed with the ART-P telescope in 1990-1992 and more than 100 
type I X-ray bursts from these sources were observed. For each of the
sources we investigated in detail the recurrence times between bursts,
the bursts time profiles and their dependence on the bursts flux, the
spectral evolution of source emission in the persistent state and during
bursts. Two bursters (SLX1732-304, 4U1724-307) located in the
globular clusters Terzan 1 and 2 were investigated also using the RXTE 
data.
}

%
%

\vspace{10ex}

\contribution
%
%
{On the pulsed optical emission from radio pulsars}
%
%
%
{\speaker{I.F.~Malov}\inst{1}}
{%
\inst{1}%
Pushchino Radio Astronomy Observatory%
\instaddr{, Pushchino, 142292, Moscow Region, Russia \\}}
{malov@prao.psn.ru}
%
%
%
{}
{}
{}{}{}


\abstext
{
The formula for a radio pulsar luminosity associated 
with synchrotron emission of the primary beam is obtained. 
The formula is based on the model of an emitting torus at 
the light cylinder and on the solution of the kinetic equation for 
pitch-angle distribution of relativistic particles. The 
high correlation between the observed optical luminosity of radio 
pulsars and the parameter $\dot P / P^4$ is found: 
$\log \left( \frac{L_{opt}}{L_{Crab}} \right)$ = (1.30 $\pm$ 0.19) 
$\log{\dot P_{-14} / P^4} - 4.21 \pm 1.02$ (the correlation coefficient 
$\rho$ = 0.97 $\pm$ 0.14). Here $P$ is the pulsar period, $\dot P$ 
is its derivative. This correlation 
allows one to predict  
possible optical emission from several dozens of pulsars 
(in particularly, from all pulsars with $P < 0.1$ sec). 
Comparison of this prediction with multiwavelength
observations of radio pulsars shows that the predicted 
list contains all 27 known emitters of hard radiation. The shift 
of maximum frequency in the synchrotron spectrum to higher 
frequencies with decreasing period $P$ is predicted. This 
prediction is in agreement with data for the same 27 pulsars. 
The obtained results show that the synchrotron model describes 
the main properties of non-thermal optical and harder emission 
of radio pulsars.
}

%
%

\newpage

\contribution
%
%
{First detection of pulsed radio emission from an AXP}
%
%
%
{V.M.~Malofeev\inst{1}, \speaker{O.I.~Malov}\inst{1}}
{%
\inst{1}%
Pushchino Radio Astronomy Observatory%
\instaddr{, Pushchino, 142292, Moscow Region, Russia \\}}
{malofeev@prao.psn.ru}
%
%
%
{}
{}
{}{}{}


\abstext
{
We report on the discovery and some investigations 
of pulsed radio emission with period 6.98 s from the anomalous 
X-ray pulsar (AXP) 1E 2259+586. The observations were made from 
March 1999 to April 2001 at 111.5 MHz with the Large Phased Array 
in Pushchino. 
The mean flux density is about 70 mJy, the integrated profile is narrow 
and its width at 50 percent of maximum intensity is approximately 
120 ms. The dispersion measure is 80 $\pm$ 5 pc cm$^{-3}$ that gives 
the distance to the pulsar of about 3.6 kpc. This value is confirmed with 
the estimation of the distance to SNR G109.1$-$1.0 (3.6 -- 4.7 kpc) 
with the X-ray - bright central pulsar 1E 2259+586. 
}

%
%

\vspace{10ex}
\label{malov2}

\contribution
%
%
{X-ray spectral variability of the LMXB and \nl Z-source GX340+0}
%
%
%
{\speaker{S.V.~Molkov}\inst{1}, S.A.~Grebenev\inst{1}, A.A.~Lutovinov\inst{1}}
{%
\inst{1}%
Space Research Institute%
\instaddr{, 84/32, Profsoyuznaja st., 117810, Moscow, Russia \\}}
{molkov@hea.iki.rssi.ru}
%
%
%
{}
{}
{}{}{}


\abstext
{
We present the results of analysis of the PCA/RXTE data obtained 
during long ($\sim$ 390 ks) pointing towards the Z-source GX~340+0. 
The complete Z track has been traced out for this source in 
the color-color diagram. For the analysis we separated the whole 
set of data (24 separate observations) in 16-s time segments and 
accumulated X-ray spectra for each of the segments. We studied 
spectral behaviour of the source as a function of its position 
in the color-color diagram. In general the spectra could not be 
fitted with any simple model. We used two-component model: 
bremsstrahlung plus black-body with fixed photoelectric absorption. 
Our analysis reveals that the role of black-body emission increases 
during transition from the horizontal branch of Z track to the 
flared branch.
}

%
%

\newpage

\contribution
%
%
{Familon emissivity of magnetized plasma}
%
%
%
{\speaker{E.N.~Narynskaya}\inst{1}, N.V.~Mikheev\inst{1}}
{%
\inst{1}%
Yaroslavl State University%
\instaddr{, Sovietskaya 13, 150000, Yaroslavl, Russia \\}}
{elenan@univ.uniyar.ac.ru}
%
%
%
{}
{}
{}{}{}


\abstext
{
Emission of familons in the processes $e^- \to e^-
+ f$, $ e^- \to \mu + f$ in a magnetized plasma is
investigated in the model in which familons either have 
or have not direct coupling to leptons via plasmon.
Contributions of the lowest and excited Landau levels
are analyzed. The differential probabilities and integral familon
effect on the plasma are calculated. It is shown
that in the process $e^- \to \mu + f$ P -- odd 
interference phenomenon leads to familon force
acting on  plasma along the magnetic field.
}

%
%

\vspace{10ex}

\contribution
%
%
{Dynamic effects in a collapsar envelope due to neutrino \nl
propagation through a strongly magnetized plasma}
%
%
%
{\speaker{I.S.~Ognev}\inst{1}, A.A.~Gvozdev\inst{1}}
{%
\inst{1}%
Yaroslavl State University%
\instaddr{, Sovietskaya 13, 150000, Yaroslavl, Russia \\}}
{ognev@univ.uniyar.ac.ru}
%
%
%
{}
{}
{}{}{}


\abstext
{
Dominated neutrino-nucleon processes in a 
supernova remnant 
around a collapsing
star with a strong magnetic field are investigated. 
The asymmetry of the envelope along the field direction
produced by
neutrino momentum transfer is calculated.
It is shown that in the toroidal magnetic field these processes can
develop a torque which quickly spins up the envelope. The influence of
neutrino ``spin-up'' effect on dynamics of the collapsar envelope is
discussed and numerically estimated.
}

%
%

\newpage

\contribution
%
%
{r-process in neutron star mergers and beta-delayed fission }
%
%
%
{\speaker{I.V.~Panov}\inst{1}, F.-K.~Thielemann\inst{2}}
{%
\inst{1}%
Institute for Theoretical and Experimental Physics,%
\instaddr{ Moscow, 117259, Russia \\}
\inst{2}%
University of Basel%
\instaddr{, Klingelbergstr. 82, CH-4056 Basel, Switzerland \\}}
{panov@mpa-garching.mpg.de, panov@vitep5.itep.ru}
%
%
{http://www.itep.ru/lab230/ns2001.html}
{http:/$\!$/www.itep.ru/lab230/ns2001.html}
{}{}{}


\abstext
{The studies of beta-delayed fission in r-process have a long
history, but they have mainly been focused on cosmochronometers. 
It has been thought that its effect has small influence on the
majority of r-process products, taken into account realistic fission
barriers.
For that reason the majority of investigations of the r-process
employed
beta-delayed fission of transuranium nuclei in a very simplified manner:
$P_{\beta df} \equiv P_{sf}=1$ for all A $>$ $A_{fiss}$ 
(see [1] and references therein).

That is why reliability 
of both, beta-delayed fission rates 
(first of all, for cosmochronology) and 
mass distribution of fission products
(important mainly for formation of nuclei with mass numbers A $<$ 130
[2]), has not been studied.
In this work we make an attempt to solve these problems numerically.     
The kinetic network [3]  calculations
of r-process for conditions in neutron star mergers  
[4] have been performed. Different theoretical beta-delayed 
fission probabilities have been used, and different 
mass distributions of fission products have been considered. 

Calculations with different fission rates  show strong 
dependence on theoretical physics input. 
Our calculations  give better agreement with
observations, first of all in relative yields for peaks $A\approx
$130, 196 and for nuclei with $ Z \approx$ 44--48. 
The problem of realistic mass distribution of nuclear fission
products is still open because of poor knowledge of fission  
of very neutron rich transuranium nuclei. 
However there are some indications
from both, nuclear physics and astrophysical observations [5], 
that asymmetric fission should be the main fission mechanism. 
The masses
of fission products have to be 
determined taking into account  shell effects. 
Preliminary results show significant dependence of the yields 
of nuclei-cosmochronometers from different fission data involved in
calculations. 
Comparison of the results with observations 
of r-elements in very metal poor stars 
may help to find limits on  
probable mass distribution of fission products.
}

\references{

\item
 J.J.~Cowan, B.~Pfeiffer, Kratz K.-L. et al., ApJ 521, 194 (1999) 

\item  I.V.~Panov, C.~Freiburghaus, F.-K.~Thielemann.
In: Proc. of 10 Workshop on Nucl. Astroph., Ringberg, 2000,
eds. W.~Hillebrand, E.~M\"oller. MPA/P12, P.73

\item I.V.~Panov, S.I.~Blinnikov, F.-K.~Thielemann,
 Astronomy Letters.  27, 279 (2001) 

\item C.~Freiburghaus, S.~Rosswog, F.-K.~Thielemann.
  ApJ  525, L121 (1999) 

\item C.~Sneden, J.J.~Cowan, I.I.~Ivans et al.
 ApJ  533, L139 (2000)

}

\newpage

\contribution
%
%
{Observed parameters of microstructure \nl in pulsar radio emission}
%
%
%
{\speaker{M.V.~Popov}\inst{1}, V.I.~Kondratev\inst{1}}
{%
\inst{1}%
Astro Space Centre FIAN%
\instaddr{, 117810, Moscow, Russia \\}}
{mpopov@asc.rssi.ru}
%
%
%
{}
{}
{}{}{}


\abstext
{
The microstructure of several bright pulsars was 
investigated with a time resolution of 62.5 ns. The pulsars 
were observed with the 70-m NASA/DSN radio telescope at Tidbinbilla, 
Australia, at a frequency of 1650 MHz. Histograms of microstructure 
time scales show steep increase toward shorter time scales followed 
by a sharp cutoff at about $5-10$ $\mu$s. The shortest micropulse 
detected has a width of 2 $\mu$s.
No unresolved nanopulses or pulse structure with submicrosecond 
time scale were found. The statistics of the micropulses and 
their quasi-periodicities differ significantly for two 
components of PSR B1133+16. Microstructure quasi-periodicities 
are most likely unrelated to any modes of vibrations of neutron stars.
}

%
%

\vspace{10ex}

\contribution
%
%
{Isolated neutron stars as X-ray sources: \nl
accretion vs. cooling}
%
%
%
{\speaker{S.B.~Popov}\inst{1}, M.E.~Prokhorov\inst{1},
M.~Colpi\inst{2}, R.~Turolla\inst{3}, A.~Treves\inst{4}}
{%
\inst{1}%
Sternberg Astronomical Institute,%
\instaddr{ Universitetskij pr. 13, 119899, Moscow, Russia \\} 
\inst{2}%
University Milano-Bicocca,%
\instaddr{ Paizza della Scienza 3, 20126, Milano, Italy \\}
\inst{3}%
University of Padova,%
\instaddr{ Via Marzolo 8, 35131, Padova, Italy \\}
\inst{4}%
University of Como%
\instaddr{, Via Lucini 3, 22100, Como, Italy\\}}
{polar@sai.msu.ru}
%
%
%
{http://xray.sai.msu.ru/~polar/}
{http:/$\!$/xray.sai.msu.ru/$\sim$polar/}
{astro-ph/0009225}{astro-ph/0011564}{astro-ph/0101031}


\abstext
{
We briefly review observational appearance of
isolated neutron stars which are not observed as
normal radio pulsars.

In some details we discuss dim X-ray sources
in globular clusters. We present a simple 
population synthesis model 
of old isolated neutron stars in
globular clusters to test suggestion made by Pfahl 
and Rappaport (2000), that these objects are powered
by accretion of interstellar medium onto isolated 
neutron stars.

We discuss different interpretations (cooling vs. 
accretion) of 7 dim ROSAT sources 
("The Magnificent Seven"). Log N- Log S distribution
of accreting and cooling neutron stars is obtained.
We suggest, that most of these sources are young
cooling neutron stars. Otherwise magnetic field decay
is necessary.

Influence of investigation of neutron stars in
dim X-ray sources on nearby fields of astrophysics
is briefly mentioned.
}

%
%

\newpage

\contribution
%
%
{Evolution of isolated neutron stars in globular clusters: \nl 
number of accretors}
%
%
%
{\speaker{S.B.~Popov}\inst{1}, M.E.~Prokhorov\inst{1}}
{%
\inst{1}%
Sternberg Astronomical Institute%
\instaddr{, Universitetskij pr. 13, 119899, Moscow, Russia \\}}
{polar@sai.msu.ru}
%
%
%
{http://xray.sai.msu.ru/~polar/}
{http:/$\!$/xray.sai.msu.ru/$\sim$polar/}
{astro-ph/0102201}{}{}


\abstext
{
With a simple model from the point of view of population 
synthesis we try to verify an
interesting suggestion made by Pfahl \& Rappaport (2000) 
that dim sources in globular clusters
(GCs) can be isolated accreting neutron stars (NSs). 
Simple estimates show, that we can expect
about $0.5-1$ accreting isolated NS per typical GC with 
$M=10^5$ M$_{\odot}$ which agrees
with observations. Properties of old accreting 
isolated NSs in GCs are briefly
discussed. We suggest that accreting NSs in GCs experienced 
significant magnetic field decay. 
}

%
%

\vspace{10ex}

\contribution
%
%
{Relation of gamma-ray bursts to formation of compact stars}
%
%
%
{\speaker{K.A.~Postnov}\inst{1}}
{%
\inst{1}%
Sternberg Astronomical Institute%
\instaddr{, Universitetskij pr. 13, 119899, Moscow, Russia \\}}
{pk@sai.msu.ru}
%
%
%
{}
{}
{}{}{}


\abstext
{
We review current observational evidence
and theoretical considerations on 
formation of relativistic compacts
stars during cosmic gamma-ray bursts.
}

%
%

\newpage

\contribution
%
%
{Thermal structure and cooling of neutron stars}
%
%
%
{\speaker{A.Y.~Potekhin}\inst{1}, D.G.~Yakovlev\inst{1}}
{%
\inst{1}%
Ioffe Physical Technical Institute%
\instaddr{, Politekhnicheskaya 26, St. Petersburg 194021 \\}}
{palex@astro.ioffe.rssi.ru}
%
%
%
{http://www.ioffe.rssi.ru/astro/NSG/NSG-Pub1.html}
{http:/$\!$/www.ioffe.rssi.ru/astro/NSG/NSG-Pub1.html}
{astro-ph/9909100}{astro-ph/0105261}{}


\abstext
{
Thermal structure of neutron stars with magnetized envelopes
is studied using modern physics input -- in particular,
updated thermal conductivity of a dense magnetized plasma
typical for neutron-star envelopes [1].
The relation between the internal ($T_{\rm int}$)
and local surface temperatures is calculated
and fitted by analytic expressions
for magnetic field strengths $B$ from 0 to 10$^{16}$~G
and arbitrary inclination of the field lines to the surface.
The luminosity of a neutron star with dipole magnetic field
is calculated and fitted as a function
of $B$, $T_{\rm int}$, stellar mass and radius.

In addition, we simulate cooling
of neutron stars with magnetized envelopes,
using the modern cooling code [2].
In particular, we analyse magnetic field effects 
in the ultramagnetized envelopes
of magnetars on the cooling curves. 
Finally, we demonstrate that the magnetic field
of the Vela pulsar strongly affects 
observational constraints on the values of
critical temperatures of neutron and proton superfluids
in its core.
}

%
%
\references{

\item
A.Yu.~Potekhin, Astron. Astrophys., 351, 787 (1999)

\item
O.Yu.~Gnedin, D.G.~Yakovlev, A.Y.~Potekhin. 
MNRAS 2001, 324, 725 (\arxiv{astro-ph/0012306})

\item
A.Yu.~Potekhin, D.G.~Yakovlev, Astron.\& Astrophys., accepted (2001)

}

\vspace{10ex}

\contribution
%
%
{Distribution of NS and BH masses \nl and mechanism of 
SN explosion}
%
%
%
{\speaker{M.E.~Prokhorov}\inst{1}, K.A.~Postnov\inst{1}}
{%
\inst{1}%
Sternberg Astronomical Institute%
\instaddr{, Universitetskij pr. 13, 119899, Moscow, Russia \\}}
{mike@sai.msu.ru}
%
%
%
{}
{}
{}{}{}


\abstext
{
The observed distribution of masses of compact remnants of
 massive star evolution (neutron stars and black holes)
 is analysed and its relation with plausible core collapse
 supernova mechanisms is discussed. It is argued that
 the observed absence of compact stars with masses 1.5--3.0 $M_\odot$
 is in favor of the magnetorotational mechanism of supernova explosion
 and soft equation of state of neutron star matter with a
 limiting mass near 1.5 $M_\odot$. Observational consequences of
 this hypothesis are discussed.
}

%
%

\newpage

\contribution
%
%
{Short-term X-ray variability of neutron stars: \nl current status
and perspectives}
%
%
%
{\speaker{M.G.~Revnivtsev}\inst{1}, M.R.~Gilfanov\inst{1},
E.M.~Churazov\inst{1}, R.A.~Sunyaev\inst{1}}
{%
\inst{1}%
Space Research Institute%
\instaddr{, 84/32, Profsoyuznaja st., 117810, Moscow, Russia \\}}
{revnivtsev@hea.iki.rssi.ru}
%
%
%
{}
{}
{}{}{}


\abstext
{
The black holes and neutron stars are the fastest objects in the
Universe. From the very beginning of X-ray astronomy there was a
search for the shortest timescales in the X-ray variability of 
these compact objects. The great recent improvement in this topic was made
with the help of Rossi X-ray Timing Explorer observatory. In this talk 
I would like to describe our latest results on the high frequency continuum
variability of black holes and neutron stars. A special attention
will be paid to the different patterns of X-ray variability of black holes
and neutron stars at hundreds of Hz and to the continuum noise of
compact objects at the highest available frequencies -- up to $\sim$30 kHz. 
}

%
%

\vspace{10ex}

\contribution
%
%
{Lepton pair production by high-energy neutrino \nl
in an external electromagnetic field}
%
%
%
{\speaker{D.A.~Rumyantsev}\inst{1}, A.V.~Kuznetsov\inst{1}, N.V.~Mikheev\inst{1} }
{%
\inst{1}%
Yaroslavl State University%
\instaddr{, Sovietskaya 13, 150000, Yaroslavl, Russia \\}}
{rda@uniyar.ac.ru}
%
%
%
{}
{}
{hep-ph/0003216}{}{}


\abstext
{
The process of the lepton pair production ($e^- e^+, 
e^- \mu^+, \dots$) by a neutrino 
propagating in an external electromagnetic field is 
investigated in the framework of the standard model. 
Relatively simple exact expression for the probability 
of the process
as the one-dimensional integral is obtained; it is suitable 
for a quantitive analysis.
}

%
%

\newpage

\contribution
%
%
{Cyclotron scattering with mode switching 
in accretion \nl column of a magnetized neutron star}
%
%
%
{\speaker{A.V.~Serber}\inst{1}}
{%
\inst{1}%
Institute of Applied Physics%
\instaddr{, 46 Ulyanov st., 603600, Nizhny Novgorod, Russia \\}}
{serber@appl.sci-nnov.ru}
%
%
%
{}
{}
{}{}{}


\abstext
{
We consider scattering of cyclotron radiation 
at the first harmonic in a rarefied plasma near a neutron 
star with a dipole magnetic field. It is assumed that the
case of strongly inhomogeneous magnetic field is realized,
i.e., the size of the plasma region is large compared 
to the size of gyroresonance layer in the neutron-star 
magnetic field. This case is analyzed under the 
conditions that normal-wave polarization at the first 
harmonic is determined by vacuum birefringence and the 
efficient switching of the linearly-polarized ordinary 
and extraordinary modes takes place due to the cyclotron 
scattering at the first harmonic. The obtained solution
 of the corresponding equation of cyclotron-radiation 
transfer reveals that the cyclotron scattering with mode 
switching leads to an efficient depolarization of
radiation outgoing from an optically thick 
gyroresonance layer. 
}

%
%

\vspace{10ex}

\contribution
%
%
{Multiwavelength observations of isolated NSs: \nl
thermal emission vs nonthermal}
%
%
%
{\speaker{Yu.A.~Shibanov}\inst{1}}
{%
\inst{1}%
Ioffe Physical Technical Institute%
\instaddr{, Politekhnicheskaya 26, St. Petersburg 194021 \\}}
{shib@stella.ioffe.rssi.ru}
%
%
%
{}
{}
{}{}{}


\abstext
{
Multiwavelength observations are important tool  
to study the properties of thermal emission from 
cooling surfaces of isolated neutron stars (NSs), to 
distinguish thermal emission from nonthermal emission
of their magnetospheres, to understand the    
mechanisms of the magnetospheric radiation 
in different wave bands, and to investigate structure 
and properties of pulsar nebulae forming due to interaction of NSs 
with ambient matter. A wealth of new information on the multiwavelength 
radiation from radio pulsars as well as from radio quiet isolated NSs 
has been obtained in recent years year with the Chandra and XMM-Newton 
X-ray observatories, HST, and new generation of large ground 
based optical telescopes VLT and Subaru.  
We review recent results obtained in a wide 
frequency range, from IR, through optical, X-ray, to gamma-ray bands, 
discuss their implications, and prospects of further studies 
of isolated NSs.    
}

%
%

\newpage

\contribution
%
%
{Diffractive scintillations of PSR 0809+74 and \nl
PSR 0950+08 at low frequencies}
%
%
%
{\speaker{T.V.~Smirnova}\inst{1}}
{%
\inst{1}%
Pushchino Radio Astronomy Observatory%
\instaddr{, Pushchino, 142292, Moscow Region, Russia \\}}
{tania@prao.psn.ru}
%
%
%
{}
{}
{}{}{}


\abstext
{
Low frequency individual pulse observations were 
carried out for the pulsars PSR 0809+74 and PSR 0950+08 in the range 
from 64 to 111 MHz. Frequency - time structure of  emission was 
studied to separate internal (due to pulsar) and external 
intensity variations due to electron density irregularities 
in the interstellar plasma. We report characteristic time 
and frequency scales of these variations at low frequencies.
}

%
%

\vspace{10ex}

\contribution
%
%
{Giant pulses from radiopulsars}
%
%
%
{\speaker{V.A.~Soglasnov}\inst{1}, M.V.~Popov\inst{1}, V.I.~Kondratev\inst{1},
S.V.~Kostyuk\inst{1}}
{%
\inst{1}%
Astro Space Centre FIAN%
\instaddr{, 117810, Moscow, Russia \\}}
{vsoglasn@asc.rssi.ru}
%
%
%
{}
{}
{}{}{}


\abstext
{
There are two pulsars, where remarkable phenomenon of 
"giant pulses" is observed: the Crab pulsar and the 
millisecond pulsar B1937+21. We present new results of our high 
time resolution observations of giant pulses from these pulsars. 
The events with extremely high flux, 300 000 Jy (Crab) and 65 000 Jy 
(1937+21), were detected. Many properties of giant pulses become 
rather unexpected (for instance, a very short - nanosecond - 
duration of GP from 1937+21), they are important for understanding 
physics of neutron star magnetosphere.
}

%
%

\newpage

\contribution
%
%
{On the gamma-ray burst progenitors}
%
%
%
{\speaker{V.V.~Sokolov}\inst{1}}
{%
\inst{1}%
SAO RAS%
\instaddr{, Nizhny Arkhyz, 357147, Karachai-Cherkessia, Russia \\}}
{sokolov@sao.ru}
%
%
%
{}
{}
{}{}{}


\abstext
{
The recent observational data show that collapse of massive stars
is a more preferable scenario to produce so-called long-duration GRBs, than
naive expectations on merging of binary compact objects, which (the
NS+NS scenario) appeared before the first determination of redshifts,
measurements of spectra and images. Observed location distribution of
variable optical sources, or optical transients (OT GRB), relative to their
host galaxies allows us to suppose that GRB sources are associated with
vigorous massive star-forming in distant galaxies: spiral, irregular,
blue compact and others with transient burst of star formation in them.
At least some OT GRBs can be located directly in star-forming regions (or
in their vicinities) of these galaxies in which (massive) star formation
rate is tens times higher than in the same galaxies in the local Universe.
In addition to the characteristic ``knee'' on the light curve of OT
GRB~970508 revealed by $I$-band photometric observations with the 6-m
telescope, more evidences (GRB 970228, GRB 980326, 990712, 991208) of link
between GRBs and Type Ib/c SNe (or core-collapse SNe) were found, 
which can be
an additional argument in favor of the idea of massive stars as progenitors
of cosmic gamma-ray bursts. 
The observations of K$_{\alpha}$ lines of iron
in afterglow X-ray spectra of GRBs (970508, 970828, 991216, 000214) and the
observation of redshifted absorption feature of neutral iron (7.1 keV)
simultaneously with the GRB 990705 give additional evidence in favor
of massive stars as progenitors of GRBs.
}
%
%

\newpage

\contribution
%
%
{Evidence for a strong evolution of GRBs and \nl
constraints on the GRB intrinsic luminosity function \nl
with new GRB statistical data}
%
%
%
{\speaker{Ya.Yu.~Tikhomirova}\inst{2,3}, B.E.~Stern\inst{1,2,3}, R.~Svensson\inst{3}}
{%
\inst{1}%
Institute for Nuclear Research RAS,%
\instaddr{ Institute for Nuclear Research, 117312, Moscow, Russia \\}
\inst{2}%
Astro Space Centre FIAN,%
\instaddr{ 117810, Moscow, Russia \\}
\inst{3}%
Stockholm Observatory%
\instaddr{, SCFAB, Stockholm Observatory, Department of Astronomy, 
SE-106 91 Stockholm, Sweden \\}}
{jana@anubis.asc.rssi.ru}
%
%
%
{}
{}
{}{}{}


\abstext
{
We present new constraints on the evolution of GRBs sources 
and on their 
intrinsic luminosity function obtained with the recent data: the uniform
sample of 3300 long GRBs found in the BATSE continuous records and  the 
sample of 17 GRBs with measured redshifts. The latter sample has three GRBs
with a very high intrinsic peak photon flux. If we fix the bright end 
of intrinsic  luminosity distribution to these events and allow the rest of 
the distribution to vary as a broken power law we find: (i) nonevolving
population models predict too large number of apparently strong GRBs, they
are rejected at the 10$^{-5}$ level; (ii) the decline of the GRBs population
from $z=1.5$ towards the present epoch is approximately as sharp as the decline 
of the star production; (iii) the intrinsic luminosity function
behaves as $dN/dP \sim P^{-1.5}$ through at least two orders of magnitude. 
The evolution of gamma bursters at large redshifts cannot be constrained. 
The data fit is not sensitive to the cosmological parameters.
}

%
%

\newpage

\contribution
%
%
{Some features of weak and strong pulses of pulsars}
%
%
%
{\speaker{M.Yu.~Timofeev}\inst{1}}
{%
\inst{1}%
Astro Space Centre FIAN%
\instaddr{, 117810, Moscow, Russia \\}}
{timofeev@anubis.asc.rssi.ru}
%
%
%
{}
{}
{}{}{}


\abstext
{
This presentation is based on the observations, which have been 
made in Effelsberg with the 100-meter radio telescope
in April 1994. Recording was produced in the mode of
the single pulses registration of five pulsars: PSR
0823+26, PSR 0950+08, PSR 1706$-$16, PSR 1929+10, PSR
2021+51 at 1.7 GHz with different time resolutions.

The goal of the programme is the investigation of some
features of individual pulses. Parameter $q$ was calculated
for any single pulse. To do this, the maximum of
intensity inside of pulse ($I_m$) was found for any
pulse and it was normalized by r.m.s. calculated
outside the pulse. Single radio pulses were divided into
groups according to the values of $q$, and thus average
profiles of total intensity in the chosen region were calculated.

It is discovered that the width of an average profile of
total intensity emission decreases while $q$ increases.
Also we have obtained some time shifts between average profiles
with different $q$. Then Gaussian fits were made for
any pulsar profile and the time shift and width
parameters were refined.

The explanation of these effects can be the following. 
Pulsar radiation is generated in a 
magnetospheric layer. Strong pulses are generated at the inner
side of the layer, close to the pulsar surface, 
where generation of the plasma oscillations may be expected. 
As the height increases, the
synchronization condition is violated 
which
leads to the emission of weaker pulses.
}

%
%

\newpage

\contribution
%
%
{Spectral and timing  properties of neutron star and \nl
black hole systems.  Theory and observations}
%
%
%
{\speaker{L.G.~Titarchuk}\inst{1}}
{%
\inst{1}%
George Mason University and Naval Research Laboratory%
\instaddr{, Center for Earth Observing and Space
          Research, George Mason University, Fairfax, VA, 22030; \\
          US Naval Research Laboratory, Washington, DC 20375-5352  \\}}
{lev@xip.nrl.navy.mil}
%
%
%
{}
{}
{}{}{}


\abstext
{
I  present  continued development of a new theoretical 
paradigm,  the Transition Layer    Model (TLM),  
for explaining features of power density 
spectra (PDS) in accreting neutron stars and black holes.   
The prevailing model for all accreting  neutron stars has been 
that an ionized  disk extends inward in the equatorial plane  
until it encounters the 
magnetosphere at the Alfven radius, inside which  MHD effects 
govern the structure and flow.  This paradigm was developed 
originally for binary pulsars. An attempt was made to extend it 
to low-mass binary systems (LMXBs) started after discovery of QPOs by EXOSAT. 
 The attempted extension  has encountered many difficulties. In the TLM 
paradigm the disk is only  partially conducting 
and its structure inside a critical radius,  
R$_{outer}$,  is governed by fluid physics.   
In TLM a pair of fluid oscillatory  modes that 
become key features related to the observed PDS.  
The  TLM has been developed through scrutiny 
of the full range of observational 
knowledge available concerning both the energy spectra and 
the PDS of neutron star and black hole sources.  
Successes include detailed fits to  the variation of 6 PDS 
features in four  LMXB sources with at most one free parameter, 
and even that one constrained. 
It is important to understand the limits of success of such a  
radical paradigm shift.  This talk covers continued development 
of  frontier areas of the theory as well as continuation of the 
crucial detailed  confrontation with observational data, done 
within the TLM  paradigm.  
Under this study both LMXB and black hole sources are studied.   
(It is possible that the paradigm will eventually extend even to the binary 
pulsars that were the starting point of the older paradigm).    
}

%
%

\newpage

\contribution
%
%
{The manifestations of different acceleration mechanisms \nl
of binaries containing neutron stars}
%
%
%
{\speaker{A.I.~Tsygan}\inst{1}, V.D.~Palshin\inst{1}}
{%
\inst{1}%
Ioffe Physical Technical Institute%
\instaddr{, Politekhnicheskaya 26, St. Petersburg 194021 \\}}
{tsygan@astro.ioffe.rssi.ru, val@pop.ioffe.rssi.ru}
%
%
%
{}
{}
{}{}{}


\abstext
{
We present the results of calculations of the distribution function
of low-mass binary systems containing a neutron star
over the orbital angular momentum orientation
for different heights $z$ above the Galactic plane.
We consider two acceleration mechanisms of the binaries:
1) acceleration by their own X-ray radiation
at the stage of intense accretion of matter onto a neutron star
with asymmetric magnetic field~[1],
2) acceleration due to a supernova explosion.

In the first case the binaries are accelerated by asymmetric X-ray radiation.
The radiative reaction force $F=\xi L_X/c$ depends on
X-ray luminosity of the neutron star, $L_X$, and on the X-ray asymmetry 
parameter $\xi$. The force points along the spin axis.
For the disk accretion onto the star, its rotational axis
aligns quickly perpendicular to the accretion disk (and the orbital plane),
and the radiative reaction force will accelerate the binary along its orbital
angular momentum. 
Thus, the systems with orbital angular momentum parallel to
the Galactic axis will tend to be at large Galactic heights
while the systems with orbital angular momentum perpendicular
the Galactic axis will be accumulated in the Galactic plane.
This is in contrast to the case of symmetric supernova explosion.
For asymmetric one, the orbital angular momentum 
orientation depends mainly on the ratio of kick velocity to 
orbital velocity of neutron star.

We use the following system parameters: 
$m_1=3M_\odot$, the pre-supernova mass;
$m_2=1M_\odot$, the companion mass; 
$m_3=1.4M_\odot$, the neutron star mass;
$P=1$~day, the orbital period 
(the relative orbital velocity $v=290$~km~s$^{-1}$).
The distribution of the kick velocities ${\bf v}_k$ 
is assumed to be Maxwellian:
$(2\pi \kappa^2)^{-3/2} \exp [-u_k^2/2\kappa^2]$, 
where ${\bf u}_k\equiv \bf{v}_k/v$ is
the dimensionless kick velocity, and $\kappa$ is the dispersion of $u_k$.

To obtain the distribution functions for different $z$
we have numerically solved the equations of motion of binaries
in the Galactic potential.
At the initial moment of time the systems 
are assumed to be in the Galactic plane,
on the orbit of the Sun.

The computations show that in case of radiative 
acceleration (for $\xi=0.1,\,0.2$)
there is a linear increase of cosine of angle between the orbital angular
momentum and the Galactic axis with increasing $z$.
In case of asymmetric supernova explosion (for $\kappa=0.2$)
the distributions of the survived binary systems over 
directions of orbital angular
momentum (for different $z$) is nearly isotropic.

The results may be valid for binary millisecond radiopulsars
if they passed the stage of intense accretion.
The possibility of pulsars recycling (to produce the millisecond pulsars)
due to disk accretion in binary systems was first considered in~[2].
}

%
%
\references{

\item
V.D.~Pal'shin, A.I.~Tsygan. Astronomy Letters 24, 131 (1998)

\item
G.S.~Bisnovatyi-Kogan, B.V.~Komberg. Pis'ma v Astron Zhurn. 2, 338 (1976)
}

\newpage

\contribution
%
%
{Thermal relaxation in young and old neutron stars}
{\speaker{D.G.~Yakovlev}\inst{1}, O.Y.~Gnedin\inst{2}, A.Y.~Potekhin\inst{1}}
{%
\inst{1}%
Ioffe Physical Technical Institute%
\instaddr{, Politekhnicheskaya 26, St. Petersburg 194021 \\}
\inst{2}%
Institute of Astronomy%
\instaddr{, Madingley Road, Cambridge CB3 0HA, England \\}}
{yak@astro.ioffe.rssi.ru}
{http://www.ioffe.rssi.ru/astro/NSG/NSG-Pub1.html}
{http:/$\!$/www.ioffe.rssi.ru/astro/NSG/NSG-Pub1.html}
{astro-ph/0012306}{}{}

\abstext
{  
Thermal relaxation in young isolated neutron stars
and in soft X-ray transients is analyzed. 

The relaxation process in young neutron stars
lasts from 10 to 100 years (Lattimer et al.\ 1994,
Gnedin et al.\ 2001). During the relaxation
stage the effective surface temperature
of the star is fairly insensitive to physical
conditions in the stellar core, being mainly  
determined by poorly known physics of matter
of subnuclear density in the neutron star crust. 
The end of the relaxation stage  
is manifested by a drop of the surface temperature, while
subsequent thermal evolution of the star is mainly
controlled by properties of matter in the stellar
core. The temporal evolution of the surface
temperature during the relaxation epoch 
carries important information on neutrino
processes, thermal conductivity, heat capacity
and superfluidity of neutrons in the stellar crust.

Another example of the thermal relaxation is provided
by soft X-ray transients.
These sources undergo periods of intense accretion
separated by long phases of quiescence. 
Their thermal state is thought to be determined
(Brown et al.\ 1998)
by energy release (mainly due to pycnonuclear
reactions) in accreted matter (Haensel and Zdunik 1990) 
sinking in deep
layers of the stellar crust under the weight of newly
accreted material. The quasisteady thermal state is
reached (Colpi et al.\ 2001)
in about $5 \times 10^4$ yrs after the transient activity
onset. The star becomes warm, some fraction of thermal
energy flows into the core and is radiated away by
neutrinos. The effective surface temperature
in the quiescence periods becomes sensitive to
neutrino emission mechanisms and superfluidity
of matter in the stellar core, and therefore, to the
stellar mass. This gives a new method (Colpi et al.\ 2001) to measure
masses of neutron stars and superfluid transition temperatures
in stellar cores.

}

%
%
\references{

\item
Brown E.F., Bildsten L., Rutledge R.E., 1998, ApJ 504, L95
(\arxiv{astro-ph/9807179})

\item
Colpi M., Geppert U., Page D., Possenti A., 2001,
ApJ 548, L175 (\arxiv{astro-ph/0010572})

\item
Gnedin O.Y., Yakovlev D.G., Potekhin A.Y., 2001,
MNRAS, 324, 725 (\arxiv{astro-ph/0012306})

\item
Haensel P., Zdunik J.L., 1990, A\&A, 227, 431; 229, 117

\item
Lattimer J.M., Van Riper K.A., Prakash M., Prakash M., 1994,
ApJ 425, 802

}

\newpage

\contribution
%
%
{From white dwarfs to black holes \nl
(70th anniversary of the theory of compact objects)}
%
%
%
{\speaker{A.F.~Zakharov}\inst{1}}
{%
\inst{1}%
Institute for Theoretical and Experimental Physics%
\instaddr{, 117259, Moscow, Russia \\}}
{zakharov@vitep5.itep.ru}
%
%
%
{}
{}
{}{}{}


\abstext
{
We discuss basic ideas which laid foundation of the black hole
concept. The major goal of the historical part is to
explain the very long way of the birth of the black hole concept.
The black hole solution was derived by K.~Schwarzschild
in 1916, but black hole concept was introduced by J.A.~Wheeler
only in 1967. We emphasize the great contribution of S.~Chandrasekhar
into development of this concept.
We discuss the basic notations of the black hole theory
and observational (astronomical) manifestations of black holes,
for example, we analyse a possibility to interpret 
the very peculiar distortion of the iron K$_{\alpha}$-line 
in such a way.
}

%
%

\vspace{1ex}

\contribution
%
%
{Accreting strange stars}
%
%
%
{\speaker{J.L. Zdunik}\inst{1}}
{%
\inst{1}%
N.~Copernicus Astronomical Center, Polish Academy of Sciences%
\instaddr{, Bartycka 18, PL-00-716 Warszawa, Poland \\}}
{jlz@camk.edu.pl}
%
%
%
{}
{}
{astro-ph/0002394}{astro-ph/0104116}{}


\abstext
{
The models of the rotating strange stars (SS) with crust
are presented in the context of the spin up by accretion.
Calculations are performed within the framework of general
relativity. Equation of state of strange quark matter is based
on the MIT Bag Model with massive strange quarks and lowest order
QCD interactions. The crust is described by the BPS
equation of state.
 
The properties of the rotating strange stars and main differences with
respect to neutron stars are discussed (the large oblateness of rotating SS,
the gravitational field in the outer spacetime, the properties of the 
innermost stable circular orbit [1,2,3]).

The evolutionary tracks for the strange stars accreting the matter from the
margi\-nally stable orbit are presented. 
If all the particle angular momentum is transferred to the star
the evolutionary tracks terminate at
mass-shed limit for almost all initial values of the stellar mass.  Only 
stars with the mass very close to the maximum one can terminate by
encountering the radial instability. 
However the collapse is the most probable fate of the
accreting SS provided that only  50\% of the particle 
angular momentum is deposited onto the star.  
The maximum amount of the accreted mass is about 0.6$\,M_{\odot}$, the 
required mass to spin up star to millisecond periods is 
$\sim$  0.2$\,M_{\odot}$.
}

%
%
\references{

\item
Zdunik J.L.,  P. Haensel, D. Rosinska, E. Gourgoulhon, 2000, A\&A 356, 612

\item
Zdunik J.L, Gourgoulhon E., 2001, Phys Rev D, 63, 087501

\item
Zdunik J.L, Haensel, P., Gourgoulhon E., 2001,  A\&A, 
in press (\arxiv{astro-ph/0104116})

}


\newpage
\thispagestyle{empty}


\markboth{\sl Contents}
         {\sl Physics of Neutron Stars -- 2001}

\tableofcontents

\end{document}